\documentclass[12pt]{article}

\usepackage{eqsection,subeqnarray,indent,amsfonts,amssymb}
\usepackage{bm}    %% For bold math symbols
\usepackage{cite}  %% For writing intervals of citations as xx--yy
\usepackage{graphicx}
\usepackage{eepic} %% For defining plot symbols \fancyplus etc

\begin{document}

\bibliographystyle{plain}

\date{February 12, 2004}

\title{Dynamic Critical Behavior of the \break
       Swendsen--Wang Algorithm \break
       for the Three-Dimensional Ising Model
      }
\author{
  {\small Giovanni Ossola}                  \\[-0.2cm]
  {\small Alan D.~Sokal}                  \\[-0.2cm]
  {\small\it Department of Physics}       \\[-0.2cm]
  {\small\it New York University}         \\[-0.2cm]
  {\small\it 4 Washington Place}          \\[-0.2cm]
  {\small\it New York, NY 10003 USA}      \\[-0.2cm]
  {\small\tt GIOVANNI.OSSOLA@PHYSICS.NYU.EDU},
                    {\small\tt SOKAL@NYU.EDU}   \\[-0.2cm]
  {\protect\makebox[5in]{\quad}}  % To force authors' names to be written
                                  %   vertically, one above another.
                                  % (\author seems to put them side-by-side
                                  %   if there is room.)
  \\
}
\vspace{0.5cm}

\maketitle
\thispagestyle{empty}   % Suppress page number on front page.

%\ltapprox and \gtapprox produce > and < signs with twiddle underneath
\def\spose#1{\hbox to 0pt{#1\hss}}
\def\ltapprox{\mathrel{\spose{\lower 3pt\hbox{$\mathchar"218$}}
 \raise 2.0pt\hbox{$\mathchar"13C$}}}
\def\gtapprox{\mathrel{\spose{\lower 3pt\hbox{$\mathchar"218$}}
 \raise 2.0pt\hbox{$\mathchar"13E$}}}
\def\inapprox{\mathrel{\spose{\lower 3pt\hbox{$\mathchar"218$}}
 \raise 2.0pt\hbox{$\mathchar"232$}}}

%%\doublespace

\begin{abstract}
We have performed a high-precision Monte Carlo study of the dynamic
critical behavior of the Swendsen--Wang algorithm for the three-dimensional
Ising model at the critical point.
For the dynamic critical exponents associated to the
integrated autocorrelation times of the ``energy-like'' observables,
we find $z_{{\rm int},{\cal N}} = z_{{\rm int},{\cal E}}
   = z_{{\rm int},{\cal E}'}  = 0.459 \pm 0.005 \pm 0.025$,
where the first error bar represents statistical error
(68\% confidence interval)
and the second error bar
represents possible systematic error due to corrections to scaling
(68\% subjective confidence interval).
For the ``susceptibility-like'' observables,
we find $z_{{\rm int},{\cal M}^2} = z_{{\rm int},{\cal S}_2} =
   0.443 \pm 0.005 \pm 0.030$.
For the dynamic critical exponent associated to the
exponential autocorrelation time,
we find $z_{\rm exp} \approx 0.481$.
Our data are consistent with the Coddington--Baillie conjecture
$z_{\rm SW} = \beta/\nu \approx 0.5183$,
especially if it is interpreted as referring to $z_{\rm exp}$.
\end{abstract}

\bigskip
\noindent
{\bf PACS codes:}  05.50.+q, 05.10.Ln, 64.60.Cn, 64.60.Ht.

\bigskip
\noindent
{\bf Key Words:}  Ising model;  Potts model;
Swendsen--Wang algorithm; cluster algorithm; Monte Carlo;
autocorrelation time; dynamic critical exponent.

\clearpage

\newtheorem{defin}{Definition}[section]
\newtheorem{definition}[defin]{Definition}
\newtheorem{prop}[defin]{Proposition}
\newtheorem{proposition}[defin]{Proposition}
\newtheorem{lem}[defin]{Lemma}
\newtheorem{lemma}[defin]{Lemma}
\newtheorem{guess}[defin]{Conjecture}
\newtheorem{ques}[defin]{Question}
\newtheorem{question}[defin]{Question}
\newtheorem{prob}[defin]{Problem}
\newtheorem{problem}[defin]{Problem}
\newtheorem{thm}[defin]{Theorem}
\newtheorem{theorem}[defin]{Theorem}
\newtheorem{cor}[defin]{Corollary}
\newtheorem{corollary}[defin]{Corollary}
\newtheorem{conj}[defin]{Conjecture}
\newtheorem{conjecture}[defin]{Conjecture}

\newtheorem{pro}{Problem}
\newtheorem{clm}{Claim}
\newtheorem{con}{Conjecture}

%
% For examples (numbered by sections, separately from theorems etc)
% Note that the \label does not include the section number;
%   that has to be put in by hand in the \ref.
%
\newcounter{example}[section]
\newenvironment{example}%
{\refstepcounter{example}
 \bigskip\par\noindent{\bf Example \thesection.\arabic{example}.}\quad
}%
%{\hbox{\kern1pt\vrule height6pt width4pt depth1pt\kern1pt}}
%{\hbox{\hskip 6pt\vrule width6pt height7pt depth1pt \hskip1pt}}
{\quad $\Box$}
\def\bexam{\begin{example}}
\def\eexam{\end{example}}

\renewcommand{\theenumi}{\alph{enumi}}
\renewcommand{\labelenumi}{(\theenumi)}
\def\prf{\par\noindent{\bf Proof.\enspace}\rm}
\def\rmk{\par\medskip\noindent{\bf Remark.\enspace}\rm}

\newcommand{\be}{\begin{equation}}
\newcommand{\ee}{\end{equation}}
\newcommand{\bmath}{\begin{displaymath}}
\newcommand{\emath}{\end{displaymath}}
\newcommand{\<}{\langle}
\renewcommand{\>}{\rangle}
\newcommand{\widebar}{\overline}
\def\reff#1{(\protect\ref{#1})}
\def\spose#1{\hbox to 0pt{#1\hss}}
\def\ltapprox{\mathrel{\spose{\lower 3pt\hbox{$\mathchar"218$}}
 \raise 2.0pt\hbox{$\mathchar"13C$}}}
\def\gtapprox{\mathrel{\spose{\lower 3pt\hbox{$\mathchar"218$}}
 \raise 2.0pt\hbox{$\mathchar"13E$}}}
\def\textprime{${}^\prime$}
\def\proof{\par\medskip\noindent{\sc Proof.\ }}
\newcommand{\qed}{\quad $\Box$ \medskip \medskip}
\def\proofof#1{\bigskip\noindent{\sc Proof of #1.\ }}
\def\half{ {1 \over 2} }
\def\third{ {1 \over 3} }
\def\twothird{ {2 \over 3} }
\def\smfrac#1#2{\textstyle{#1\over #2}}
\def\smhalf{ \smfrac{1}{2} }
\newcommand{\real}{\mathop{\rm Re}\nolimits}
\renewcommand{\Re}{\mathop{\rm Re}\nolimits}
\newcommand{\imag}{\mathop{\rm Im}\nolimits}
\renewcommand{\Im}{\mathop{\rm Im}\nolimits}
\newcommand{\sgn}{\mathop{\rm sgn}\nolimits}
\newcommand{\var}{\mathop{\rm var}\nolimits}
\newcommand{\cov}{\mathop{\rm cov}\nolimits}
\def\hboxscript#1{ {\hbox{\scriptsize\em #1}} }

\newcommand{\restrict}{\upharpoonright}
\renewcommand{\emptyset}{\varnothing}

\def\Z{{\mathbb Z}}
\def\ZZ{{\mathbb Z}}
\def\R{{\mathbb R}}
\def\C{{\mathbb C}}
\def\CC{{\mathbb C}}
\def\N{{\mathbb N}}
\def\NN{{\mathbb N}}
\def\Q{{\mathbb Q}}

\newcommand{\scra}{{\mathcal{A}}}
\newcommand{\scrb}{{\mathcal{B}}}
\newcommand{\scrc}{{\mathcal{C}}}
\newcommand{\scre}{{\mathcal{E}}}
\newcommand{\scrf}{{\mathcal{F}}}
\newcommand{\scrg}{{\mathcal{G}}}
\newcommand{\scrh}{{\mathcal{H}}}
\newcommand{\scrl}{{\mathcal{L}}}
\newcommand{\scrm}{{\mathcal{M}}}
\newcommand{\scrn}{{\mathcal{N}}}
\newcommand{\scro}{{\mathcal{O}}}
\newcommand{\scrp}{{\mathcal{P}}}
\newcommand{\scrr}{{\mathcal{R}}}
\newcommand{\scrs}{{\mathcal{S}}}
\newcommand{\scrt}{{\mathcal{T}}}
\newcommand{\scrv}{{\mathcal{V}}}
\newcommand{\scrw}{{\mathcal{W}}}
\newcommand{\scrz}{{\mathcal{Z}}}
\newcommand{\scrbt}{{\mathcal{BT}}}
\newcommand{\scrbf}{{\mathcal{BF}}}

\newcommand{\bgamma}{\boldsymbol{\gamma}}
\newcommand{\bsigma}{\boldsymbol{\sigma}}
\renewcommand{\pmod}[1]{\;({\rm mod}\:#1)}

% Array for subscripts

\newenvironment{sarray}{
	  \textfont0=\scriptfont0
	  \scriptfont0=\scriptscriptfont0
	  \textfont1=\scriptfont1
	  \scriptfont1=\scriptscriptfont1
	  \textfont2=\scriptfont2
	  \scriptfont2=\scriptscriptfont2
	  \textfont3=\scriptfont3
	  \scriptfont3=\scriptscriptfont3
	\renewcommand{\arraystretch}{0.7}
	\begin{array}{l}}{\end{array}}

\newenvironment{scarray}{
	  \textfont0=\scriptfont0
	  \scriptfont0=\scriptscriptfont0
	  \textfont1=\scriptfont1
	  \scriptfont1=\scriptscriptfont1
	  \textfont2=\scriptfont2
	  \scriptfont2=\scriptscriptfont2
	  \textfont3=\scriptfont3
	  \scriptfont3=\scriptscriptfont3
	\renewcommand{\arraystretch}{0.7}
	\begin{array}{c}}{\end{array}}

\setlength{\unitlength}{0.13cm}
\newsavebox{\fancyplusb}
\newsavebox{\fancycrossb}
\newsavebox{\fancysquareb}
\newsavebox{\fancydiamondb}

\savebox{\fancyplusb}(2,2){
\thinlines
\put(0,-1){\line(0,1){2}}
\put(-1,-0.5){\line(0,1){1}}
\put(-1,0){\line(1,0){2}}
\put(1,-0.5){\line(0,1){1}}
\put(-0.5,-1){\line(1,0){1}}
\put(-0.5,1){\line(1,0){1}}
}

\savebox{\fancycrossb}(2,2){
\thinlines
\put(-1,-1){\line(1,1){2}}
\put(-1,1){\line(1,-1){2}}
\put(-1,-0.5){\line(1,-1){0.5}}
\put(0.5,1){\line(1,-1){0.5}}
\put(-1,0.5){\line(1,1){0.5}}
\put(0.5,-1){\line(1,1){0.5}}
}

\savebox{\fancysquareb}(2,2){
\thinlines
\put(-1,-1){\line(1,1){0.5}}
\put(-1,1){\line(1,-1){0.5}}
\put(0.5,0.5){\line(1,1){0.5}}
\put(0.5,-0.5){\line(1,-1){0.5}}
\put(-0.5,-0.5){\line(1,0){1}}
\put(0.5,-0.5){\line(0,1){1}}
\put(-0.5,-0.5){\line(0,1){1}}
\put(-0.5,0.5){\line(1,0){1}}
}

\savebox{\fancydiamondb}(2,2){
\thinlines
\put(-0.7,0){\line(1,1){0.7}}
\put(0,0.7){\line(1,-1){0.7}}
\put(-0.7,0){\line(1,-1){0.7}}
\put(0,-0.7){\line(1,1){0.7}}
\put(-1.2,0){\line(1,0){0.5}}
\put(0.7,0){\line(1,0){0.5}}
\put(0,0.7){\line(0,1){0.5}}
\put(0,-1.2){\line(0,1){0.5}}
}

\newcommand{\fancyplus}{\begin{picture}(2,2)
\usebox{\fancyplusb}
\end{picture}}

\newcommand{\fancycross}{\begin{picture}(2,2)
\usebox{\fancycrossb}
\end{picture}}

\newcommand{\fancysquare}{\begin{picture}(2,2)
\usebox{\fancysquareb}
\end{picture}}

\newcommand{\fancydiamond}{\begin{picture}(2,2)
\usebox{\fancydiamondb}
\end{picture}}

%%%%%%%%%%%%% BEGINNING OF THE TEXT %%%%%%%%%%%%%%%%%%%%%%%%%%%%%%%%%%%%

\section{Introduction}  \label{sec_intro}

% {\bf Rewrite all, emphasizing 3D Ising, but still putting it in
%    broader context of Potts models as a function of $q$ and $d$!!!
%    Play down Li--Sokal bound, mention work on it in 2D 2,3,4-state Potts,
%    but emphasize that in 3D Ising the bound is clearly NOT sharp!!!
%    Main thing is to review confusing nature of past work.}

Monte Carlo (MC) simulations
\cite{Binder_79,Binder_84,Binder_92,Sokal_Cargese,Newman_99,Landau_00,Binder_02}
have become a standard and powerful tool for gaining
nonperturbative insights into
statistical-mechanical systems and lattice field theories.
However, their practical success is severely limited by critical
slowing-down: the autocorrelation time $\tau$ --- that is, roughly speaking,
the time needed to produce one ``statistically independent'' configuration ---
diverges near a critical point.
More precisely, for a finite system of linear size $L$ at criticality,
we expect a behavior $\tau \sim L^{z}$ for large $L$.
The power $z$ is a {\em dynamic critical exponent}\/,
and it depends on both the system and the Monte Carlo algorithm.

Local Monte Carlo algorithms (such as single-site Metropolis or heat bath)
generally have a dynamic critical exponent $z \gtapprox 2$.
This makes it very hard to get high-precision data very close to the
critical point on large lattices.

In some cases, a much better dynamical behavior can be obtained by including
non-local moves, such as cluster flips.\footnote{
   See \cite{Sokal_Cargese,Janke_98} for reviews of
   collective-mode Monte Carlo methods.
}
In particular, the Swendsen--Wang (SW) cluster algorithm \cite{Swendsen_87}
for the ferromagnetic $q$-state Potts model achieves
a significant reduction in $z$ compared to the local algorithms:
one has $z$ between 0 and 1,
where the exact value depends on $q$ and on the dimensionality of the lattice.
The most favorable case is the two-dimensional (2D) Ising model ($q=2$),
for which the best currently available numerical estimate is
$z = 0.222 \pm 0.007$ \cite{Salas_Sokal_Ising_v1}
(see also the discussion below).
In other cases, the performance of the SW algorithm
is somewhat less impressive but still quite good:
e.g., $z = 0.514 \pm 0.006$ for the 2D 3-state Potts model
\cite{Salas_Sokal_Potts3},
$z \approx 1$ for the 2D 4-state Potts model
\cite{Salas_Sokal_AT,Salas_Sokal_Potts4},
$z \approx 0.45$--0.75 for the 3D Ising model
\cite{Swendsen_87,Wolff_89,Wang_90,Coddington_92,%
Kerler_93a,Kerler_93b,Hennecke_93,Wang_02},
and $z \approx 0.86$--1  for the 4D Ising model
\cite{Klein_89,Ray_89,Coddington_92,Persky_96,Jerrum_in_prep}.
Clearly, we would like to understand why the SW algorithm works so well
in some cases and less well in others.
We hope in this way to obtain new insights into the
dynamics of non-local Monte Carlo algorithms,
with the ultimate aim of devising new and more efficient algorithms.

There is at present no adequate theory for predicting the
dynamic critical behavior of an SW-type algorithm.
However, Li and Sokal \cite{Li_Sokal} have proven that the
autocorrelation times of the Swendsen--Wang algorithm
for the $q$-state Potts ferromagnet
are {\em bounded below}\/ by a multiple of the specific heat:
\be
  \tau_{{\rm int},{\cal N}}, \;
  \tau_{{\rm int},{\cal E}}, \;
  \tau_{{\rm int},{\cal E}'}, \;
  \tau_{\rm exp}  \;\ge\; {\rm const} \times C_H   \;.
\label{Li_Sokal_bound_CH}
\ee
Here ${\cal N}$ is the bond occupation in the SW algorithm,
${\cal E}$ is the energy, ${\cal E}'$ is the nearest-neighbor connectivity,
and $C_H$ is the specific heat;
$\tau_{\rm int}$ and $\tau_{\rm exp}$ denote the integrated and
exponential autocorrelation times, respectively \cite{Sokal_Cargese}.
As a result one has for the dynamic critical exponents
\be
  z_{{\rm int},{\cal N}}, \;
  z_{{\rm int},{\cal E}}, \;
  z_{{\rm int},{\cal E'}}, \;
  z_{\rm exp}  \;\ge\; \alpha/\nu \;,
\label{Li_Sokal_bound}
\ee
where $\alpha$ and $\nu$ are the standard {\em static}\/ critical exponents.
Thus, the SW algorithm cannot {\em completely}\/ eliminate
the critical slowing-down
if the specific heat is divergent at criticality.\footnote{
   The bound \reff{Li_Sokal_bound_CH}/\reff{Li_Sokal_bound} has also been
   proven to hold \cite{Salas_Sokal_AT}
   for the direct SW-type algorithm \cite{Wiseman_Domany}
   for the Ashkin-Teller (AT) model \cite{Ashkin_Teller,Baxter},
   which interpolates between the 2-state (Ising) and 4-state Potts models.
}
The physical mechanism underlying the Li--Sokal proof
is the slow evolution of the bond occupation ${\cal N}$:
its mean-square change per SW iteration is of order $V$ (volume),
while its variance in the equilibrium distribution is of order $V C_H$;
this leads to \reff{Li_Sokal_bound_CH}.
In addition, Salas and Sokal \cite{Salas_Sokal_Potts3} have proven,
under a very mild and eminently plausible condition\footnote{
   The condition is that the normalized autocorrelation functions
   of the bond occupation at time lags 1 and 2,
   i.e.\  $\rho_{\cal NN}(1)$ and $\rho_{\cal NN}(2)$,
   should be bounded away from zero as the lattice size $L$
   tends to infinity.
},
that all three ``energy-like'' observables have the same
dynamic critical exponent:
\be
   z_{{\rm int},{\cal N}}   \;=\;
   z_{{\rm int},{\cal E}}   \;=\;
   z_{{\rm int},{\cal E}'}    \;.
 \label{equal4}
\ee
For details on all these results, see \cite[Section 2.2]{Salas_Sokal_Potts3}.

One important question is whether the Li--Sokal bound
\reff{Li_Sokal_bound_CH}/\reff{Li_Sokal_bound} is sharp or not.
An affirmative answer would imply that we could use the bound to predict
the dynamic critical exponent(s) $z$ given only the static
critical exponents of the system.
There are three possibilities:
\begin{enumerate}
\item[i)]
The bound \reff{Li_Sokal_bound_CH} is {\em sharp}\/
(i.e., the ratio $\tau/C_H$ is bounded), so that \reff{Li_Sokal_bound} is an
{\em equality}\/.

\item[ii)]
The bound is {\em sharp modulo a logarithm}\/
(i.e., $\tau/C_H \sim \log^p L$ for $p>0$).

\item[iii)]
The bound is {\em not sharp}\/
(i.e., $\tau/C_H \sim L^p$ for $p>0$), so that \reff{Li_Sokal_bound} is a
{\em strict inequality}\/.
\end{enumerate}
The best currently available data concern the two-dimensional Potts models
with $q=2,3,4$, for which we have
\begin{center}
\begin{tabular}{lll}
   $q=2$:  &  $\alpha/\nu = 0 \, (\times\log)$  &
      $z_{{\rm int},{\cal E}'} = 0.222 \pm 0.007$ \cite{Salas_Sokal_Ising_v1}
   \\[2mm]
   $q=3$:  &  $\alpha/\nu = 2/5$  &
      $z_{{\rm int},{\cal E}'} = 0.514 \pm 0.006$ \cite{Salas_Sokal_Potts3}
   \\[2mm]
   $q=4$:  &  $\alpha/\nu = 1 \, (\times\log^{-3/2})$  &
      $z_{{\rm int},{\cal E}\hphantom{'}} =
                                 0.876 \pm 0.011$ \cite{Salas_Sokal_Potts4}
\end{tabular}
\end{center}
Here the values of $\alpha/\nu$ are exact
\cite{Nienhuis_84,DiFrancesco_97,Salas_Sokal_Potts4},
while the values of $z$ are the best available numerical estimates
from pure power-law fits.
Note, however, that the estimate of $z$ for $q=4$ {\em cannot}\/
be correct, as it violates the Li--Sokal bound \reff{Li_Sokal_bound};
presumably it is corrupted by the same multiplicative logarithmic
corrections that afflict the specific heat.\footnote{
   Indeed, a pure power-law fit to the Monte Carlo data for the specific heat
   yielded $\alpha/\nu = 0.770 \pm 0.008$ \cite{Salas_Sokal_Potts4}.
}
For this reason, the papers
\cite{Salas_Sokal_Ising_v1,Salas_Sokal_Potts3,Salas_Sokal_Potts4}
analyzed also the ratio $\tau/C_H$
in order to test directly the sharpness of the Li--Sokal bound.
It was found that the data for $q=2,3,4$ are consistent with two scenarios:
either the Li--Sokal bound is non-sharp by a very small power
($p \approx 0.06$--0.12),
or else it is sharp modulo a logarithm (possibly with power $p=1$).
Not surprisingly, it is exceedingly difficult to distinguish numerically
between these two scenarios.

For the three- and four-dimensional Ising models, by contrast,
the Li--Sokal bound \reff{Li_Sokal_bound} is clearly {\em not}\/ sharp:
the numerical estimates of $z$ are much larger than $\alpha/\nu$.
It follows that another physical mechanism,
beyond the one captured in the Li--Sokal proof,
must be principally responsible for the critical slowing-down
in these cases;
but it is far from clear what this mechanism is.
A natural first step towards identifying this mechanism would be
to obtain accurate numerical estimates for $z$ in the
three- and four-dimensional Ising models.
Unfortunately, the numerical results in the four-dimensional case
are almost nonexistent\footnote{
   The only numerical study of which we are aware is
   \cite{Coddington_92},
   which yielded $z_{{\rm int},{\cal E}} = 0.86 \pm 0.02$,
   based on lattices of size $4 \le L \le 16$
   (which by present-day standards are much too small).
   In addition, there have been numerical \cite{Ray_89,Persky_96}
   and analytic \cite{Ray_89,Persky_96,Jerrum_in_prep}
   studies of the Swendsen--Wang algorithm for the Ising ferromagnet
   on the complete graph
   (also known as the Curie--Weiss or ``mean-field'' model),
   which indicate $z=1$.
   This model is {\em presumed}\/ to lie in the same dynamic
   universality class as the Ising model on a regular lattice
   of dimension $d \ge 4$, but high-precision numerical tests
   of this quite plausible conjecture are lacking.
},
and in the three-dimensional case are wildly contradictory.
In chronological order, the results for the three-dimensional
Ising model are as follows:
\begin{quote}
\begin{itemize}
   \item[1987:]  $z_{{\rm exp},{\cal E}} = 0.75 \pm 0.01$, based on
      unspecified lattice sizes \cite{Swendsen_87}
   \item[1989:]  $z_{{\rm int},{\cal E}} = z_{{\rm int},{\cal M}^2}
                  = 0.50 \pm 0.03$, based on $L \le 64$ \cite{Wolff_89}
   \item[1990:]  $z_{{\rm exp},{\cal C}_1} = 0.46 \pm 0.01$
      (${\cal C}_1 =$ size of the largest cluster),
      based on $L \le 89$ extrapolated to $L=\infty$ \cite{Wang_90}\footnote{
         Wang \cite{Wang_90} also studied the magnetization (including sign)
         of the largest cluster in a variant of the Swendsen--Wang algorithm
         in which the largest cluster is not flipped.
         But it is far from clear whether this observable corresponds
         to {\em any }\/observable in the standard Swendsen--Wang algorithm.
         Indeed, Wang found that the exponential autocorrelation time
         of this observable is several times {\em larger}\/ than that
         of ${\cal C}_1$.  And since there is good reason to believe
         (see Section~\ref{sec_data_analysis_tauexp} below)
         that ${\cal C}_1$ does indeed have a significant overlap
         with the slowest mode in the Swendsen--Wang algorithm,
         this suggests that Wang's observable is {\em not}\/ interpretable
         within the standard Swendsen--Wang algorithm, but rather
         represents a {\em new}\/ slow mode in the variant algorithm.
}
   \item[1992:]  $z_{{\rm int},{\cal E}} = 0.54 \pm 0.02$,
      based on $L \le 48$ \cite{Coddington_92}
   \item[1993:]  $z_{\rm exp} = 0.61 \pm 0.02$,
          $z_{{\rm int},{\cal E}} = 0.58 \pm 0.02$,
          $z_{{\rm int},{\cal M}^2} = 0.57 \pm 0.02$ and
          $z_{{\rm int}, |{\cal M}|} = 0.55 \pm 0.02$,
      based on $L \le 32$ \cite{Kerler_93b}
   \item[1993:]  $z_{{\rm int},{\cal E}} = 0.49 \pm 0.02$ and
          $z_{{\rm int}, |{\cal M}|} = 0.48 \pm 0.01$,
      based on $L \le 60$ \cite{Hennecke_93}
   \item[2002:]  $z_{{\rm int},{\cal E}} \approx 0.50$,
      based on $L \le 128$ \cite{Wang_02}\footnote{
         A pure power-law fit to the raw data of Wang, Kozan and Swendsen
         \cite{Wang_02} yields a decent $\chi^2$ if (and only if)
         $L_{\rm min} \ge 32$.
         Our preferred fit is $L_{\rm min} = 32$, and yields
         $z_{{\rm int},{\cal E}} = 0.502 \pm 0.012$
         ($\chi^2 = 0.440$, 1 DF, level = 50.7\%).
         We thank Jian-Sheng Wang for supplying us with these raw data.
}
\end{itemize}
\end{quote}
The reason for the discrepancies is unclear,
but there does seem to be a tendency for the
estimates of $z$ to decrease as larger lattices are used ---
an effect that could easily be understood as arising from
corrections to scaling.
   
The purpose of this paper is to restudy the dynamic critical behavior
of the SW algorithm for the three-dimensional Ising model,
using much larger lattices (up to $L = 256$),
vastly higher statistics
(well over $10^7$ SW iterations at each lattice size),
and a careful finite-size-scaling analysis.
For the dynamic critical exponents associated to the
integrated autocorrelation times of the ``energy-like'' observables,
we find
\be 
   z_{{\rm int},{\cal N}} \;=\; z_{{\rm int},{\cal E}}
   \;=\; z_{{\rm int},{\cal E}'}  \;=\; 0.459 \pm 0.005 \pm 0.025
   \;,
\ee
where the first error bar represents statistical error
(68\% confidence interval)
and the second error bar
represents possible systematic error due to corrections to scaling
(68\% subjective confidence interval).
For the ``susceptibility-like'' observables,
we find 
\be
   z_{{\rm int},{\cal M}^2}  \;=\;  z_{{\rm int},{\cal S}_2}  \;=\; 
   0.443 \pm 0.005 \pm 0.030
   \;.
\ee
Finally, for the dynamic critical exponent associated to the
exponential autocorrelation time, we obtain the rough estimate
\be
   z_{\rm exp}  \;\approx\;  0.481
   \;.
\ee
It is possible that some or all of these exponents are in fact exactly equal.

The present paper is organized as follows:
Section~\ref{sec_setup} reviews the basics of the Swendsen--Wang algorithm
and the definitions of autocorrelation times and observables.
In Section~\ref{sec_statistical} we discuss our methods of
statistical data analysis.
In Section~\ref{sec_simul} we summarize our Monte Carlo simulations.
In Section~\ref{sec_data_analysis} we present the analysis of our
dynamic data.
In Section~\ref{sec_discussion} we discuss our results
in the light of various conjectures that have been made by previous workers.
The static data from our simulations will be analyzed
in a separate paper \cite{sw3d_ising_static}.

\section{Basic set-up and notation} \label{sec_setup}

\subsection{Potts model and Swendsen--Wang algorithm} \label{sec_setup_notation}

The $q$-state Potts model assigns to each lattice site $i$ a spin variable
$\sigma_i$ taking values in the set $\{1,2,\ldots,q\}$;
these spins interact through the reduced Hamiltonian
\be
{\cal H}_{\rm Potts}   \;=\;   - \beta \sum_{\< ij \>}
      (\delta_{\sigma_i,\sigma_j} - 1) \; ,
\label{Potts_Hamiltonian}
\ee
where the sum runs over all the nearest-neighbor pairs $\<ij\>$
(each pair counted once).
To simplify the notation we shall henceforth write
$\delta_{\sigma_i,\sigma_j} \equiv \delta_{\sigma_b}$ for a bond $b=\<ij\>$.
The ferromagnetic case corresponds to $\beta \ge 0$.
The partition function is defined as
\be
Z   \;=\;   \sum_{\{\sigma\}} e^{-{\cal H}}   \;=\;
    \sum_{\{\sigma\}}
       \exp\!\left[ \beta \sum\limits_b (\delta_{\sigma_b} - 1) \right]
    \;.
\label{Partition_function}
\ee
Finally, the Boltzmann weight of a configuration $\{\sigma\}$ is given by
\be
W_{\rm Potts}(\{\sigma\})   \;=\;
 {1 \over Z} \, \exp\!\left[ \beta \sum\limits_b (\delta_{\sigma_b} - 1) \right]
   \;=\;   {1 \over Z} \prod_b ( 1-p + p \delta_{\sigma_b} )
\label{Potts_weight}
\ee
where $p=1-e^{-\beta}$.

The idea behind the Swendsen--Wang (SW) algorithm
\cite{Swendsen_87,Edwards_Sokal} is to decompose the
Boltzmann weight \reff{Potts_weight} by introducing
new dynamical variables $n_b=0,1$ living on the bonds of the lattice,
and to simulate the joint model of old and new variables
by alternately updating one set of variables conditional on the other set.
The Boltzmann weight of the joint model is
\be
W_{\rm joint}(\{\sigma\},\{n\})   \;=\;   {1 \over Z} \prod_b \left[
   (1-p) \delta_{n_b,0} + p \delta_{\sigma_b} \delta_{n_b,1} \right]
   \;.
\label{joint_weight}
\ee
The marginal distribution of
\reff{joint_weight} with respect to the spin variables reproduces
the Potts-model Boltzmann weight \reff{Potts_weight}.
The marginal distribution of \reff{joint_weight} with
respect to the bond variables is the
Fortuin--Kasteleyn \cite{Kasteleyn_69,Fortuin_Kasteleyn_72,Fortuin_72}
random-cluster model with parameter $q$:
\be
W_{\rm RC}(\{n\})   \;=\;
   {1 \over Z} \left[ \prod_{b \colon\; n_b=1} p \right]
 \left[ \prod_{b \colon\; n_b=0} (1-p) \right] q^{{\cal C}(\{n\})}
\label{RC_weight}
\ee
where ${\cal C}(\{n\})$ is the number of connected components (including
one-site components) in the graph whose edges are the bonds with $n_b=1$.

We can also consider the conditional probabilities of the joint
distribution \reff{joint_weight}.
The conditional distribution of the $\{n\}$ given the
$\{\sigma\}$ is as follows: independently for each bond $b = \<ij\>$, one sets
$n_b=0$ when $\sigma_i \neq \sigma_j$, and sets
$n_b=0$ and 1 with probabilities $1-p$ and $p$ when $\sigma_i=\sigma_j$.
Finally, the conditional distribution of the $\{\sigma\}$ given the
$\{n\}$ is as follows: independently for each connected cluster, one sets
all the spins $\sigma_i$ in that cluster equal to the same value, chosen
with uniform probability from the set $\{1,2,\ldots,q\}$.

The Swendsen--Wang algorithm simulates the
joint probability distribution \reff{joint_weight}
by alternately applying the two conditional distributions just described.
That is, we first erase the current $\{n\}$ configuration,
and generate a new $\{n\}$ configuration from the
conditional distribution given $\{\sigma\}$;
we then erase the current $\{\sigma\}$ configuration,
and generate a new $\{\sigma\}$ configuration from the
conditional distribution given $\{n\}$.
A single step of the SW algorithm consists of these two ``half-steps''.

\subsection{Autocorrelation functions and autocorrelation times}

Let ${\cal O}$ be any observable
(i.e.\ any function of $\{\sigma\}$ and $\{n\}$),
and let ${\cal O}(t)$ be its evolution in Monte Carlo time
(where one unit of time corresponds to a single step of the
Swendsen--Wang algorithm).
The unnormalized autocorrelation function associated to the
observable ${\cal O}$ is defined as
\be
   C_{\cal OO}(t) \;\equiv\; \< {\cal O}(s) {\cal O}(s+t) \> - \<{\cal O}\>^2
   \;,
\label{def_c}
\ee
where the expectations are taken {\em in equilibrium}\/.
The corresponding normalized autocorrelation function is defined as
\be
  \rho_{\cal OO}(t) \;\equiv\;
      {C_{\cal OO}(t) \over C_{\cal OO}(0) }  \;=\;
      {C_{\cal OO}(t) \over \var({\cal O}) }  \;.
\label{def_rho}
\ee
The integrated and exponential autocorrelation times
associated to the observable ${\cal O}$ are defined as\footnote{
   For a general Markov chain,
   the ``$\lim$'' in \reff{def_tau_exp} should strictly speaking
   be replaced by ``$\limsup$'',
   and $\rho_{\cal OO}(t)$ should be replaced by its absolute value.
   But in the Swendsen--Wang algorithm it can be {\em proven}\/
   that the limit really exists,
   and that $\rho_{\cal OO}(t) \ge 0$ for all $t$;
   this follows from the spectral representation
   \cite{Li_Sokal} \cite[Section 2.2]{Salas_Sokal_Potts3}.
}
\begin{eqnarray}
\tau_{{\rm int},{\cal O}}  &=& {1 \over 2} \sum_{t=-\infty}^{\infty}
   \rho_{\cal OO}(t)
  \label{def_tau_int} \\[2mm]
\tau_{{\rm exp},{\cal O}}  &=& \lim_{|t|\rightarrow\infty}
   {-|t| \over \log \rho_{\cal OO}(t) }
  \label{def_tau_exp}
\end{eqnarray}
Finally, the exponential autocorrelation time of the system is defined as
\be
   \tau_{\rm exp}  \;=\;  \sup_{{\cal O}}  \tau_{{\rm exp},{\cal O}}  \;,
\ee
where the supremum is taken over all observables ${\cal O}$.
This autocorrelation time measures the decay rate of the
slowest mode of the system.
All observables that are not orthogonal to this slowest mode
satisfy $\tau_{{\rm exp},{\cal O}} = \tau_{\rm exp}$.

It is important to remember that there is not just one autocorrelation time,
but many:  namely, $\tau_{\rm exp}$ as well as $\tau_{{\rm int},{\cal O}}$
for each observable ${\cal O}$.
In all but the most trivial Markov chains,
these autocorrelation times are {\em not}\/ equal.
Correspondingly, there are many dynamic critical exponents:
namely, $z_{\rm exp}$ as well as $z_{{\rm int},{\cal O}}$
for each observable ${\cal O}$.
These exponents {\em may}\/ in some cases be equal
(i.e., the corresponding autocorrelation times may scale proportionally
 as the critical point is approached),
but they need not be;
this is a detailed dynamical question,
and the answer will vary from model to model.

\subsection{Observables to be measured} \label{sec_obs}

As just explained, the Swendsen--Wang algorithm is most naturally
defined in the general context of the $q$-state Potts ferromagnet.
It is therefore most convenient and natural to use a formalism
that is valid for arbitrary $q$; 
at the end we can specialize to the Ising case $q=2$.

The nicest ``geometric'' representation of Potts spins is
the {\em hypertetrahedral representation}\/, defined as follows:
Let $\{ {\bf e}^{(\alpha)} \} _{\alpha=1}^q$
be unit vectors in $\R^{q-1}$ satisfying
${\bf e}^{(\alpha)} \cdot {\bf e}^{(\beta)} =
 (q \delta^{\alpha\beta} - 1)/(q-1)$.
Geometrically, these vectors point from the center to the vertices
of a unit hypertetrahedron in $\R^{q-1}$.
We then represent a Potts spin $\sigma_x \in \{1,2,\ldots,q\}$
by the unit vector $\bsigma_x \equiv {\bf e}^{(\sigma_x)}$ in $\R^{q-1}$.
This representation captures the $S_q$ (permutation group) symmetry
of the Potts Hamiltonian \reff{Potts_Hamiltonian},
and for $q=2$ it reduces to the usual representation
of Ising spins $\bsigma_x = \pm 1$.
We have in particular
\be
   \bsigma_x \cdot \bsigma_y  \;=\;
   {q \delta_{\sigma_x,\sigma_y} - 1  \over  q-1}   \;,
\ee
so that the Potts Hamiltonian can be written equivalently as
\begin{subeqnarray}
    \scrh  & = &
        - \beta_{\rm Potts} \, \sum_{\< ij \>} \delta_{\sigma_i,\sigma_j}
               \;+\; {\rm const}   \\[1mm]
     & = &
        - \beta_{\rm tetr} \, \sum_{\< ij \>} \bsigma_i \cdot \bsigma_j
               \;+\; {\rm const}
\end{subeqnarray}
where
\be
   \beta_{\rm tetr}   \;=\;  {q-1 \over q} \, \beta_{\rm Potts}  \;.
\ee
For $q=2$ this yields $\beta_{\rm Ising} = \beta_{\rm Potts} / 2$,
where $\beta_{\rm Ising} \equiv \beta_{\rm tetr}$
corresponds to the usual Ising normalization for the inverse temperature.

Let us now consider the $q$-state Potts ferromagnet on a
$d$-dimensional periodic hypercubic lattice of linear size $L$.
We write $V = L^d$ for the number of sites,
and $B = dL^d$ for the number of bonds.
We shall consider the following observables:
\begin{itemize}
   \item (minus) the total energy
\begin{subeqnarray}
{\cal E} &\equiv& \sum_{\<xy\>} \bsigma_x \cdot \bsigma_y \\
         &=& \sum_{\<xy\>} \frac{q \delta_{\sigma_x,\sigma_y} - 1}{q-1}
\label{def_energy}
\end{subeqnarray}
where the sum runs over all the nearest-neighbor pairs $\<xy\>$
(each pair counted once).
   \item the bond occupation
\be
 {\cal N} \;\equiv\; \sum_{\<xy\>} n_{xy}
\label{def_n}
\ee
   \item the nearest-neighbor connectivity
       (which is an energy-like observable \cite{Salas_Sokal_Potts3})
\be
{\cal E}' \;\equiv\; \sum_{\<xy\>} \gamma_{xy} \;,
\label{def_energyp}
\ee
where $\gamma_{xy}$ equals 1 if both ends of the bond $\<xy\>$ belong to the
same cluster, and 0 otherwise.
More generally, the connectivity $\gamma_{ij}$ can be defined
for an arbitrary pair $i,j$ of sites:
\be
  \gamma_{ij}(\{n\}) \;=\; \left\{ \begin{array}{ll}
       1 & \quad \hbox{\rm if $i$ is connected to $j$} \\
       0 & \quad \hbox{\rm if $i$ is not connected to $j$}
       \end{array} \right.
\ee
We shall also use higher connectivities, such as
\be
  \gamma_{ijkl}(\{n\}) \;=\; \left\{ \begin{array}{ll}
       1 & \quad \hbox{\rm if $i,j,k,l$ are all connected together} \\
       0 & \quad \hbox{\rm otherwise}
       \end{array} \right.
\ee
   \item the squared magnetization
\begin{subeqnarray}
{\cal M}^2  &=& \left( \sum_x \bsigma_x \right)^2        \\
            &=& {q \over q-1} \sum_{\alpha=1}^q \left(
     \sum_x \delta_{\sigma_x,\alpha} \right)^2 - {V^2 \over q-1}
 \label{def_msquare}
\end{subeqnarray}
%% where $\sigma_x \equiv {\bf e}^{(\sigma_x)} \in \R^{q-1}$
%% is the Potts spin in the hypertetrahedral representation
%% and $V$ is the number of lattice sites.
%
   \item higher powers of the magnetization
\be
   {\cal M}^{2n}   \;=\;  ({\cal M}^2)^n
\ee
(In this paper, we measured only ${\cal M}^2$ and ${\cal M}^4$.)
   \item the square of the Fourier transform of the spin variable
at the smallest allowed non-zero momentum
\begin{subeqnarray}
{\cal F}    &=&
{1 \over d} \sum_{j=1}^d \left| \sum_x \bsigma_x \, e^{2\pi i x_j/L} \right|^2
     \\
  &=& {q \over q-1} \times {1 \over d} \sum_{j=1}^d \sum_{\alpha=1}^q
   \left| \sum_x \delta_{\sigma_x,\alpha} \, e^{2\pi i x_j/L} \right|^2
 \label{def_f}
\end{subeqnarray}
where $(x_1,x_2,\ldots,x_d)$  are the Cartesian coordinates of point $x$.
Note that ${\cal F}$ is normalized to be comparable to
its zero-momentum analogue ${\cal M}^2$.
  \item  the number of clusters (= connected components)
     and the mean-square and mean-fourth-power size of the clusters
\begin{eqnarray}
{\cal C}  \;=\; {\cal S}_0  &=& \sum_{C} 1
        \label{def_s0}   \\[2mm]
{\cal S}_2  &=& \sum_{C} \#(C)^2  \;=\;  \sum_{x,y} \gamma_{xy}
        \label{def_s2}   \\[2mm]
{\cal S}_4  &=& \sum_{C} \#(C)^4  \;=\;  \sum_{x,y,u,v} \gamma_{xyuv}
        \label{def_s4}
\end{eqnarray}
where the sum runs over all the clusters $C$ of activated bonds,
and $\#(C)$ is the number of sites in the cluster $C$.
  \item  the size $\scrc_i$ of the $i$th largest cluster
         ($\scrc_1 \ge \scrc_2 \ge \scrc_3 \ge \ldots$).
         In this work we measured only $\scrc_1$, $\scrc_2$ and $\scrc_3$.
\end{itemize}

\noindent
{}From these observables we compute the following expectation values:
\begin{itemize}
   \item (minus) the energy density $E$ per bond
\be
E  \;=\; {1 \over B} \< {\cal E} \>   \;,
\label{def_energy_density}
\ee
where $B=dV$ is the number of bonds in the lattice,
so that a perfectly disordered (resp.\ ferromagnetically ordered) state has $E=0$ (resp.\ $E=1$)
   \item the specific heat per bond
\be
C_H \;=\; {1 \over B} \var({\cal E})  \equiv {1\over B} \left[
         \< {\cal E}^2 \> - \< {\cal E} \>^2 \right]
\label{def_cv}
\ee
   \item the magnetic susceptibility
\be
  \chi \;=\; {1 \over V} \< {\cal M}^2 \>
\label{def_susceptibility}
\ee
   \item  the correlation function at momentum $(2\pi/L,0,\ldots,0)$
\be
 F \;=\; {1 \over V} \< {\cal F} \>
 \label{def_f_density}
\ee
   \item the second-moment correlation length
\be
  \xi \;=\; {1 \over 2 \sin(\pi/L)} \left( {\chi \over F} - 1 \right)^{1/2}
  \label{def_xi_sec3}
\ee
%%  This definition of the correlation length is {\em not}\/ equal
%%  to the exponential
%%  correlation length (=1/mass gap); but it is expected that both correlation
%%  lengths scale in the same way as we approach the critical point.
%
   \item the mean number of clusters
\be
   S_0  \;=\;  \< \scrs_0 \>
\ee
   \item the mean size of the $i$th largest cluster
\be
   C_i  \;=\;  \< \scrc_i \>
\ee
\end{itemize}

For each observable ${\cal O}$ discussed above,
we have measured its autocorrelation function $\rho_{\cal OO}(t)$
and have used this to estimate
the corresponding integrated autocorrelation time
$\tau_{{\rm int},{\cal O}}$.
In Section~\ref{sec_statistical} we explain
how we derived estimates of the mean values and the error bars
for both static and dynamic quantities.

\bigskip

\noindent
{\bf Remarks.}
1.  Using the Fortuin--Kasteleyn identities
\cite{Kasteleyn_69,Fortuin_Kasteleyn_72,Fortuin_72,Sokal_Cargese},
which arise from the formulae for conditional expectations in
the joint measure \reff{joint_weight},
it is not difficult to show that
\begin{eqnarray}
\<  n_{xy} \>  & = &  p \, \< \delta_{\sigma_x,\sigma_y}  \>
       \\[2mm]
\<  \bsigma_x \cdot \bsigma_y  \> &=&  \< \gamma_{xy}  \>
       \\[2mm]
\<  \bsigma_x \cdot \bsigma_y  \;\,  \bsigma_u \cdot \bsigma_v \> &=&
\Bigl\< \gamma_{xy} \gamma_{uv} + {1 \over q-1} \left( \gamma_{xu} \gamma_{yv} +
                 \gamma_{xv} \gamma_{yu} - 2 \gamma_{xyuv} \right) \Bigr\>
\end{eqnarray}
and hence that
\begin{eqnarray}
\< {\cal N} \>  & = &  p \, {(q-1) \< {\cal E} \> + B  \over q}
     \label{check_NE}   \\[2mm]
\< {\cal E} \>  & = &  \< {\cal E}' \>  \label{check_EE'}   \\[2mm]
\< {\cal M}^2 \>    & = &   \< {\cal S}_2 \>   \label{check_stwo}  \\[2mm]
\< {\cal M}^4 \>    & = &   {q+1\over q-1} \< {\cal S}_2^2 \> -
                               {2  \over q-1} \< {\cal S}_4 \>
                                                  \label{check_sfour}
\end{eqnarray}
where $p = 1 - e^{-\beta_{\rm Potts}}$
is the Swendsen--Wang bond probability
and $B=dL^d$ is the number of bonds in the lattice.
As a check on the correctness of our simulations,
we have tested these identities to high precision,
in the following way:
Instead of comparing directly the left and right sides of each equation,
which are strongly positively correlated in the Monte Carlo simulation,
a more sensitive test
is to define new observables corresponding to the differences
(i.e., ${\cal E} - {\cal E}'$ and so forth).
Each such observable should have mean zero,
and the error bars on the sample mean
can be estimated using the standard error analysis outlined
in Section~\ref{sec_statistical} below.
The comparison to zero yields the following $\chi^2$ values:
\begin{eqnarray}
  \hbox{For \reff{check_NE}:} & \; &
     \chi^2 = 160.95 \hbox{ (182 DF, level = 87\%)}   \\
  \hbox{For \reff{check_EE'}:} & \; &
     \chi^2 = 152.63 \hbox{ (182 DF, level = 94\%)}   \\
  \hbox{For \reff{check_stwo}:} & \; &
     \chi^2 = 188.28  \hbox{ (182 DF, level = 36\%)}   \\
  \hbox{For \reff{check_sfour}:} & \; &
     \chi^2 = 190.47  \hbox{ (182 DF, level = 32\%)}
\end{eqnarray}
Here we have treated each independent run (see Section \ref{sec_simul})
as a separate data point;
DF means the number of degrees of freedom
(i.e.\ the number of independent runs),
and ``level'' means the confidence level of the fit
(defined at the beginning of Section~\ref{sec_data_analysis} below).
The agreement is excellent.

2.  As a further check on the correctness of our simulations,
we have computed both sides of the identity
\be
\rho_{\cal NN}(1) \;=\;
  1 - { q^2\ (1-p)\ E \over (q-1)^2\ p\ C_H + q^2\ (1-p)\ E }
 \label{check_rho}
\ee
proven in \cite[equation 7]{Li_Sokal}
and \cite[equation (2.16)]{Salas_Sokal_Potts3}).\footnote{
   Unfortunately, three different normalizations of the energy
   are used in \cite{Li_Sokal}, in \cite{Salas_Sokal_Potts3},
   and in the present paper.
% {\tt Updated footnote:
% We have three definitions of energy:
% \begin{itemize}
% \item Li-Sokal[89] used
%    $E = (1/V)\<\sum_{\<xy\>}\delta_{\sigma_x,\sigma_y}\> = E_1$
% \item Sokal-Salas[99] used
% $E = (1/V)\<\sum_{\<xy\>} (1-\delta_{\sigma_x,\sigma_y}\> = E_2 = d - E_1$,
% where $d$ is the number of dimensions of the lattice.
% \item Here we are using  $E = (1/B)\< \sum_{\<xy\>}
% \frac{q \delta_{\sigma_x,\sigma_y} - 1}{q-1}\> = E_3$, where $B = dV$.
% The relations with the other two definitions are
% \bmath
% E_3 = 1 - {q  \over d (q-1)} E_2 = {q \over d (q-1)} E_1 - {1 \over (q-1)}
% \emath
% For example, if we want to compare a 2-dimensional $q=2$ Potts model,
% with the analysis of Sokal-Salas[99], we fix $d=2$, $q=2$ to get
% $E_3 = 1 - E_2$ (this is the comparison that I'm doing for the Machta project,
% in which we are using the same normalization that we are using here!)
% \end{itemize}
% }
}
This is a highly nontrivial test, as it relates static quantities
(energy and specific heat) to a dynamic quantity (autocorrelation function
of the bond occupation at time lag 1). We have also checked with great
accuracy the identities \cite{Salas_Sokal_Potts3}
\begin{eqnarray}
C_{\cal EE}(t)      &=&
   {1 \over p^2} \left({q-1 \over q}\right)^2 C_{\cal NN}(t+1) \\[2mm]
\rho_{\cal EE}(t)   &=& {\rho_{\cal NN}(t+1) \over \rho_{\cal NN}(1)} \\[2mm]
C_{\cal E'E'}(t)    &=& C_{\cal EE}(t+1) \\[2mm]
\rho_{\cal E'E'}(t) &=& {\rho_{\cal EE}(t+1) \over \rho_{\cal EE}(1)}
\end{eqnarray}
%
%% {\tt GO: I changed the normalization!!!!!!}

\section{Statistical methods} \label{sec_statistical}

In this paper we are aiming at extremely high precision
for both static and dynamic quantities;
and furthermore we need to disentangle the effects of statistical errors
from the effects of systematic errors due to corrections to scaling.
For this, it is essential to obtain accurate estimates
not only of the static and dynamic quantities of interest,
but also of their {\em error bars}\/:
in this way we will be able (see Section \ref{sec_data_analysis})
to perform $\chi^2$ tests that provide an objective measure
of the goodness of fit in each scaling Ansatz.

In this section we review briefly how we performed the
statistical analysis of our raw Monte Carlo data.
In particular, we describe how to compute the estimators
for the mean value and the variance of both static and dynamic
quantities. These methods are based on well-known
results of time-series analysis \cite{Anderson,Priestley}.
More details on the methods used here can be found in
\cite[Appendix C]{Madras_Sokal}, \cite[Section 3]{Sokal_Cargese}
and \cite[Section 4]{Salas_Sokal_Potts3}.

% Then (Section \ref{sec_statistical_bunches})
% we describe an alternative analysis method,
% based on independent ``bunches'',
% and report the results of detailed cross-checks that confirm
% (with one slight exception) the validity and reliability
% of the standard time-series-analysis method.

%% \subsection{``Standard'' time-series-analysis method}
%% \label{sec_statistical_standard}

Let us consider a generic observable ${\cal O}$, whose mean is equal to
$\mu_{\cal O}$. Its corresponding unnormalized and normalized autocorrelation
functions are denoted by
$C_{\cal OO}(t) \equiv \<{\cal O}(0){\cal O}(t)\> - \<{\cal O}\>^2$ and
$\rho_{\cal OO}(t) \equiv C_{\cal OO}(t)/C_{\cal OO}(0)$,
respectively.
We also define the integrated autocorrelation time
\be
\tau_{{\rm int},{\cal O}} \;=\; {1\over 2}
                    \sum_{t=-\infty}^\infty \rho_{\cal OO}(t)  \;.
\ee

Given a sequence of $n$ Monte Carlo measurements
of the observable ${\cal O}$ --- call them
$\{ {\cal O}_1, \ldots, {\cal O}_n \}$ --- the natural estimator of the
mean $\mu_{\cal O}$ is the sample mean
\be
\overline{{\cal O}}  \;\equiv\;  {1 \over n} \sum_{i=1}^n {\cal O}_i \;.
\ee
This estimator is unbiased and has a variance
\begin{subeqnarray}
{\rm var}(\overline{{\cal O}}) &=& {1 \over n^2}
                            \sum_{r,s=1}^n C_{\cal OO}(r-s)
    \\
                              &=& {1 \over n} \sum_{t=-(n-1)}^{n-1}
    \left( 1 - {|t| \over n} \right) C_{\cal OO}(t)  \\
                              &\approx& {1 \over n}\
                              2 \tau_{ {\rm int},{\cal O} } \
                               C_{\cal OO}(0) \quad \mbox{for} \quad
    n \gg \tau_{ {\rm int},{\cal O} }
\label{var_mean}
\end{subeqnarray}
This means that the variance is a factor $2 \tau_{ {\rm int},{\cal O} }$
larger than it would be if the measurements were uncorrelated. It is, therefore,
very important to estimate the autocorrelation time $\tau_{{\rm int},{\cal O}}$
in order to ensure a correct determination of the error bar on
the (static) quantity $\mu_{\cal O}$.

The natural estimator for the unnormalized autocorrelation function
$C_{\cal OO}(t)$ is
\be
\widehat{C}_{\cal OO}(t) \;\equiv\; {1 \over n - |t|} \sum_{i=1}^{n-|t|}
           ({\cal O}_i - \mu_{\cal O})({\cal O}_{i+|t|} - \mu_{\cal O})
\ee
if the mean $\mu_{\cal O}$ is known, and
\be
\widehat{\widehat{C}}_{\cal OO}(t)
           \;\equiv\; {1 \over n - |t|} \sum_{i=1}^{n-|t|}
           ({\cal O}_i - \overline{\cal O})
           ({\cal O}_{i+|t|} - \overline{\cal O})
\ee
if the mean $\mu_{\cal O}$ is unknown.
We emphasize that, for each $t$, the estimators
$\widehat{C}_{\cal OO}(t)$ and
$\widehat{\widehat{C}}_{\cal OO}(t)$ are
{\em random variables}\/ [in contrast to $C_{\cal OO}(t)$, which is a
{\em number}\/].
The estimator $\widehat{C}_{\cal OO}(t)$ is unbiased, and
$\widehat{\widehat{C}}_{\cal OO}(t)$ is biased by terms of
order $1/n$.
The covariance matrices of $\widehat{C}_{\cal OO}$ and
$\widehat{\widehat{C}}_{\cal OO}$ are the same to leading order
in the large-$n$ limit
(i.e., $n \gg \tau_{{\rm int},{\cal O}}$), and we have
\cite{Anderson,Priestley}
\begin{eqnarray}
{\rm cov}(\widehat{C}_{\cal OO}(t),\widehat{C}_{\cal OO}(u)) &=&
{1 \over n} \sum_{i=-\infty}^{\infty} \left[ C_{\cal OO}(m) C_{\cal OO}(m+u-t)
 +C_{\cal OO}(m+u)C_{\cal OO}(m-t) \right. \nonumber \\
 & & \qquad \qquad \left. + \kappa(t,m,m+u)\right] + o\left({1\over n}\right)
 \;,
\label{cov_c}
\end{eqnarray}
where $t,u \ge 0$ and $\kappa$ is the connected 4-point autocorrelation
function
\begin{eqnarray}
\kappa(r,s,t) &\equiv&
        \< ({\cal O}_i -\mu_{\cal O})    ({\cal O}_{i+r} -\mu_{\cal O})
           ({\cal O}_{i+s} -\mu_{\cal O})({\cal O}_{i+t} -\mu_{\cal O})\>
\nonumber \\
 & & \qquad - C_{\cal OO}(r)C_{\cal OO}(t-s) - C_{\cal OO}(s)C_{\cal OO}(t-r)
 \nonumber \\
 & & \qquad
     - C_{\cal OO}(t)C_{\cal OO}(s-r) \;.
\end{eqnarray}

The natural estimator for the normalized autocorrelation function
$\rho_{\cal OO}(t)$ is
\be
\widehat{\rho}_{\cal OO}(t) \;\equiv\; {\widehat{C}_{\cal OO}(t) \over
                                  \widehat{C}_{\cal OO}(0) }
\label{estim1}
\ee
if the mean $\mu_{\cal O}$ is known, and
\be
\widehat{\widehat{\rho}}_{\cal OO}(t) \;\equiv\;
        {\widehat{\widehat{C}}_{\cal OO}(t) \over
         \widehat{\widehat{C}}_{\cal OO}(0) }
\label{estim2}
\ee
if the mean $\mu_{\cal O}$ is unknown.
The estimators $\widehat{\rho}_{\cal OO}(t)$ and
$\widehat{\widehat{\rho}}_{\cal OO}(t)$ are biased by terms of order $1/n$,
as a result of the ratios of random variables in
\reff{estim1}/\reff{estim2}.
The covariance matrices of $\widehat{\rho}_{\cal OO}$ and
$\widehat{\widehat{\rho}}_{\cal OO}$ are the same to leading order in
$1/n$.
If the process is Gaussian,
this covariance matrix is given in the large-$n$ limit by
\cite{Priestley}
\begin{eqnarray}
{\rm cov}(\widehat{\rho}_{\cal OO}(t),\widehat{\rho}_{\cal OO}(u)) &=&
{1 \over n} \sum_{m=-\infty}^{\infty} \left[
  \rho_{\cal OO}(m)\rho_{\cal OO}(m+t-u) +
  \rho_{\cal OO}(m+u)\rho_{\cal OO}(m-t)  \right. \nonumber \\
 & & + 2\rho_{\cal OO}(t)\rho_{\cal OO}(u)\rho_{\cal OO}^2(m)
     - 2\rho_{\cal OO}(t)\rho_{\cal OO}(m)\rho_{\cal OO}(m-u) \nonumber \\
 & & \left.
     - 2\rho_{\cal OO}(u)\rho_{\cal OO}(m)\rho_{\cal OO}(m-t) \right] +
     o\left({1 \over n}\right)
\label{cov_rho}
\end{eqnarray}
for $t, u \ge 0$.
If the process is {\em not}\/ Gaussian, then there are additional terms
proportional to the fourth cumulant $\kappa(m,t,t-u)$.
The simplest assumption is to consider the stochastic process to be
``not too far from Gaussian'', and drop all the terms involving $\kappa$.
If this assumption is {\em not}\/ justified, then we are introducing a bias in
the estimate of this covariance.

Finally, we shall take
the estimator for the integrated autocorrelation time to be
\cite{Madras_Sokal}
\be
\widehat{\tau}_{{\rm int},{\cal O}} \;\equiv\; {1\over 2} \sum_{t=-M}^M
            \widehat{\rho}_{\cal OO}(t)
\ee
[or the same thing with $\widehat{\widehat{\rho}}_{\cal OO}(t)$]
where $M$ is a suitably chosen number. The reason behind the cutoff $M$ is
the following: if we were to make the ``obvious'' choice $M=n+1$, then the
resulting estimator would have a variance of order 1 even in the limit
$n\rightarrow\infty$;
this is because the terms $\widehat{\rho}_{\cal OO}(t)$ with
large $t$ have errors (of order $1/n$) that {\em do not}\/ vanish as $t$
grows [cf. \reff{cov_rho}], and their number is
also large ($\sim n$).
Taking $M \ll n$ restores the good behavior of the estimator as
$n\rightarrow\infty$.
The bias
introduced by this rectangular cutoff\footnote{
   We could use more general cutoff functions, but this rectangular cutoff
   is the most convenient for the present purposes.
}
is given by
\be
{\rm bias}(\widehat{\tau}_{{\rm int},{\cal O}}) \;=\;
 - {1 \over 2} \sum_{|t| > M} \rho_{\cal OO}(t) + o\left({1\over n}\right)
\;.
\label{bias_tauint}
\ee
The variance of the estimator $\widehat{\tau}_{{\rm int},{\cal O}}$ can
be computed from the covariance \reff{cov_rho};
the final result is \cite{Madras_Sokal}
\be
{\rm var}(\widehat{\tau}_{{\rm int},{\cal O}}) \;\approx\;
{2(2M + 1) \over n} \tau_{{\rm int},{\cal O}}^2 \;,
\label{var_tauint}
\ee
where the approximation $\tau_{{\rm int},{\cal O}} \ll M \ll n$ has been made.
A good (self-consistent) choice of $M$ is the following \cite{Madras_Sokal}:
let $M$ be the smallest integer such that
$M \ge c \widehat{\tau}_{{\rm int},{\cal O}}(M)$, where $c$ is a suitable
constant.
If the normalized autocorrelation function is roughly a pure
exponential\footnote{
   This is confirmed here for the observables
   $\scrn$, $\scre$, $\scre'$, $\scrm^2$, $\scrs_2$ and $\scrc_1$
   (see Section~\ref{sec_data_analysis_tauexp}).
   Similar behavior is found in the SW algorithm for
   the two-dimensional Ising \cite{Salas_Sokal_Ising_v1},
   3-state Potts \cite{Salas_Sokal_Potts3}
   and 4-state Potts models \cite{Salas_Sokal_AT}.
},
then a choice in the range $c\approx 6$--8 is reasonable.
Indeed, if we take $\rho_{\cal OO}(t) = e^{-t/\tau}$
and minimize the mean-square error
\be
\hbox{MSE}(\widehat{\tau}_{{\rm int},{\cal O}}) \;\equiv\;
{\rm bias}(\widehat{\tau}_{{\rm int},{\cal O}})^2 +
{\rm var}(\widehat{\tau}_{{\rm int},{\cal O}})
\ee
using \reff{bias_tauint}/\reff{var_tauint}, we find that the optimal
window width is
\be
M_{\rm opt} \;=\; {\tau \over 2} \log \left( {n \over 2 \tau} \right) -1 \;.
\ee
For $n/\tau \approx 10^8$ (resp. $10^6$, $10^4$), we have
$M_{\rm opt}/\tau \approx 8.86$ (resp.\ 6.56, 4.26).
In this paper we used $c=8$ for the observables
$\scrn$, $\scre$, $\scre'$, $\scrm^2$, $\scrs_2$ and $\scrc_1$,
whose autocorrelation functions are close to a pure exponential
(see Section~\ref{sec_data_analysis_tauexp});
$c=10$ for $\scrs_0$; and $c=15$ for $\scrc_2$ and $\scrc_3$.

As noted above, we expect the estimator
$\widehat{\tau}_{{\rm int},{\cal O}}$ to have a bias of order
$\tau_{{\rm int},{\cal O}}/n$, due to the nonlinearities in
\reff{estim1}/\reff{estim2}.\footnote{
    The bias on the estimator $\widehat{\tau}_{{\rm int},{\cal O}}$ also
    induces a bias on the
    estimated variance \reff{var_mean}
    of the sample mean $\overline{\cal O}$.
    This bias is of order $1/n^2$, i.e.\ a factor $1/n$ down from
    the variance \reff{var_mean} itself.
}
To make this bias negligible we need long
runs. It has been shown empirically that this procedure works fairly well
when $n \gtapprox 10^4 \widehat{\tau}_{{\rm int},{\cal O}}$
\cite{Sokal_Cargese}.

\medskip

\noindent
{\bf Remarks.} 1. For the specific heat $C_H$ and the correlation length $\xi$,
which are ``composite'' quantities (i.e.\ not merely the mean value of
a single observable), the estimation of the error bars is a bit more
complicated.  One method is described in \cite[Section 4]{Salas_Sokal_Potts3};
a slightly better method, based on the analysis of the cross-correlation
matrix, is described in \cite{sw3d_ising_static}.

2. On most lattices we made a number of independent runs,
rather than one long run (see Section~\ref{sec_simul} below).
Our best estimate of each autocorrelation function $\rho_{\cal OO}(t)$
was then obtained by averaging the estimates
$\widehat{\widehat{\rho}}_{\cal OO}(t)$ from the individual runs,
with weights proportional to the run lengths.
Finally, the windowing procedure was performed
on the resulting best estimate of $\rho_{\cal OO}(t)$.
This is a better procedure than performing the windowing
on each run separately.

3. As a check on the correctness of the error bars produced by
our time-series-analysis method, we also used
an analysis method based on independent ``bunches''
\cite[Section 4.2]{Salas_Sokal_Potts3}.
The error bars ranged from $\approx 50\%$ to $\approx 115\%$
of those produced by the time-series-analysis method,
averaging around 80\%.
The fluctuations are not surprising,
as the bunch-method error bar has a statistical fluctuation
of order $1/\sqrt{m}$, where $m$ is the number of bunches
(in our case ranging from 10 to 35).
However, the systematic tendency toward smaller error bars
suggests that our time-series-analysis method may be
slightly overestimating the error bars,
probably due to neglect of the non-Gaussian terms in \reff{cov_rho}.
Therefore, the true statistical error bars on our raw data
and on our exponent estimates may be slightly smaller than those
reported in this paper.
This issue deserves a more detailed investigation in the future.

\section{Description of the simulations} \label{sec_simul}

We implemented the Swendsen--Wang algorithm
for the nearest-neighbor three-dimensional Ising model
on an $L \times L \times L$ simple-cubic lattice
with periodic boundary conditions.
We performed all our runs at $\beta_{\rm Ising} = 0.22165459$
(i.e.\ $\beta_{\rm Potts} = 0.44330918$),
which is Bl\"ote {\em et al.}\/'s \cite{Blote_99}
best estimate of the critical temperature
and is very near to the estimates by other workers
\cite{Ballesteros_99,Hasenbusch_99}
(see also the review \cite{Hasenbusch_01}).
We studied lattice sizes
$L=4$, 6, 8, 12, 16, 24, 32, 48, 64, 96, 128, 192, 256
and performed between $2.9 \times 10^7$ and $5 \times 10^8$ SW iterations
for each lattice size (see Table~\ref{sum_runs}).
The total data set at each $L$
corresponds to $\approx 10^6 \tau$ on the largest lattices ($L=192,256$),
at least $10^7 \tau$ on all $L \le 64$,
and nearly $10^8 \tau$ at $L=16$ (see again Table~\ref{sum_runs}).
In all cases, the statistics are high enough to permit a high accuracy in our
estimates of the static (error $\sim$ 0.01--0.12\%)
and dynamic (error $\sim$ 0.1--0.5\%) quantities.
Our results for the principal static observables are reported in
Table~\ref{table_static_full},
and for the dynamic quantities in
Tables~\ref{table_dynamic_1} and \ref{table_dynamic_2}.

The initial configuration of each run was either random or ordered,
and we discarded the first $10^5$ iterations from each run
in order to allow the system to reach equilibrium;
this discard interval is in all cases greater than
$4000 \,\tau_{{\rm int},{\cal E}'}$, which is more than sufficient.\footnote{
   Such a discard interval might seem to be much larger than necessary:
   $100 \tau_{\rm exp}$ would usually be more than enough.
   However, there is always the danger that the longest autocorrelation time
   in the system ($\tau_{\rm exp}$)
   may be much larger than the longest autocorrelation time
   that one has {\em measured}\/, because one has failed to measure
   an observable having sufficiently strong overlap with the slowest mode.
   As an undoubtedly overly conservative precaution against the possible
   (but unlikely) existence of such a (vastly) slower mode,
   we decided to discard $10^5$ iterations.
   In most cases this amounts to less than 10\% of the run,
   thus reducing the accuracy on our final estimates by less than 5\%.
   Unless there exists a vastly slower mode of which we are unaware,
   our data yield
   $\tau_{{\rm int},{\cal E}'}/\tau_{{\rm exp}} \approx 0.9$--1
   for this algorithm (see Section~\protect\ref{sec_data_analysis_tauexp}
   and Table~\ref{table_ratios_1} below).
   So the discard interval is greater than $4000 \tau_{{\rm exp}}$.
}
We checked that random and ordered initial conditions
gave identical results, within statistical error.
On some of the smaller lattices, we made a single long run
of $10^8$ iterations;  on other lattices,
we averaged the data from several (anywhere from 2 to 46) individual runs of
at least $10^6$ iterations each
(except for a small number of runs of length $5 \times 10^5$
 at $L=96, 192, 256$).
In all cases we discarded the first $10^5$ iterations of each run.
The individual runs (minus the discard) are all of length greater than
$20000 \tau_{{\rm int},{\cal E}'}$, which is long enough to allow
a good determination of the dynamic quantities.

Our program was written in Fortran 77
and run on a 1266 MHz Pentium III Tualatin processor
using the g77 Fortran compiler.
Our program requires approximately $42 L^3$ bytes memory.
The CPU time required by our program ranges from
0.39 to 0.90 $L^3$ $\mu$s/iteration,
depending on the lattice size (see Table~\ref{sum_runs}).
The sharp rise in CPU time per spin on very small lattices
arises from the ``fixed costs'' of the algorithm
(i.e.\ those that do not scale with the volume).
The slow rise in CPU time per spin on larger lattices
arises from the ``cache misses'' that occur,
due to the nonlocal nature of the Swendsen--Wang algorithm,
when the lattice no longer fits in the 512 KB cache.
The total CPU time used in these runs was approximately 17.8 years.

In the first version of our program, the random numbers were supplied
by a linear congruential generator
\be
   x_{n+1}  \;=\;  ax_n + c  \, \pmod{m}
 \label{def_LCRG}
\ee
with modulus $m=2^{48}$, increment $c=1$, and multiplier
$a=3116728$, 10430376854301, 77596615844045 or 181465474592829.
All these multipliers give good results on the spectral test in low dimensions,
compared to other multipliers for the same modulus \cite{Knuth_81,Lecuyer_99}.
We verified that the runs with the four different multipliers
gave results that are consistent within error bars
for all the major observables.
But when we analyzed the data for $\xi/L$,
which ought to behave according to the finite-size scaling Ansatz
\be
   \xi/L  \;=\;  x^\star \,+\, A L^{-\omega} \,+\, \ldots
 \label{def_FSS}
\ee
where $x^\star$ is a universal amplitude ratio
and $\omega$ is a correction-to-scaling exponent,
we found that our data fit this Ansatz very well
(with $\omega = 0.82$ from \cite{Blote_99})
{\em except for the points at $L=128$ and $L=256$}\/,
which showed deviations of magnitude 3\% ($\approx 79$ standard deviations)
and 21\% ($\approx 170$ standard deviations), respectively.
Clearly something was going very wrong!

After much work, we traced these systematic errors to the effects of
long-range correlations (at lags that are multiples of large powers of 2)
in the random-number generator \cite{Kalle_84,Filk_85,Percus_88,systematic}.
It turns out \cite{systematic} that these long-range correlations
can arise within a single bond-update half-sweep of the
Swendsen--Wang algorithm, provided that the lattice size is large enough
compared to the modulus of the random-number generator.
In a separate paper \cite{systematic} we have studied these
systematic errors in detail, in an effort to determine their approximate
magnitude as a function of the lattice size and the
random-number-generator modulus.
Suffice it to say here that the systematic errors with a
48-bit random-number generator are comparable to or larger than
our statistical errors {\em only}\/ when the lattice size is a multiple of 64,
which in this paper means $L=64,128,192,256$.
We therefore {\em discarded}\/ all the data for these lattices
($\approx\! 9.5$ years CPU time, alas!)
and performed new runs using 60-bit, 63-bit and 64-bit
random-number generators:
\begin{itemize}
    \item[{}] Modulus $m=2^{60}$, multiplier $a=454339144066433781$.
    \item[{}] Modulus $m=2^{63}$, multiplier $a=9219741426499971445$.
    \item[{}] Modulus $m=2^{64}$, multiplier $a=3202034522624059733$.
\end{itemize}
(All these multipliers give good results on the spectral test
 in low dimensions, compared to other multipliers
 for the same modulus \cite{Knuth_81,Lecuyer_99}.)
We also performed some additional runs on the smaller lattices
using these generators.
We have convinced ourselves \cite{systematic}
that generators using $\ge 60$ bits will exhibit significant systematic errors
{\em only}\/ on lattices larger than $L=256$;
in addition, they {\em may}\/ exhibit slight systematic errors,
less than about $2\sigma$, also at $L=256$
(we are currently investigating this latter issue more carefully).

We assure the reader that the data reported in the present paper
include {\em only}\/ runs using ``safe'' random-number generators,
i.e.\ $m \ge 2^{60}$ for $L=64,128,192,256$,
and $m \ge 2^{48}$ for all other $L$.
The CPU time figure of 17.8 years refers to these ``good'' runs only.

\section{Data analysis} \label{sec_data_analysis}

For each quantity ${\cal O}$, we carry out a fit to the power-law
Ansatz ${\cal O} = A L^p$
using the standard weighted least-squares method.
As a precaution against corrections to scaling,
we impose a lower cutoff $L \ge L_{\rm min}$
on the data points admitted in the fit,
and we study systematically the effects of varying $L_{\rm min}$ on the
estimates of $A$ and $p$ and on the $\chi^2$ value.
In general, our preferred fit corresponds to the smallest $L_{\rm min}$
for which the goodness of fit is reasonable
(e.g., the confidence level\footnote{
   ``Confidence level'' is the probability that $\chi^2$ would
   exceed the observed value, assuming that the underlying statistical
   model is correct.  An unusually low confidence level
   (e.g., less than 5\%) thus suggests that the underlying statistical model
   is {\em incorrect}\/ --- the most likely cause of which would be
   corrections to scaling not included in the Ansatz.
}
is $\gtapprox$ 10--20\%),
and for which subsequent increases in $L_{\rm min}$ do not cause the
$\chi^2$ to drop vastly more than one unit per degree of freedom.

The behavior of the static quantities will be discussed
in a separate paper \cite{sw3d_ising_static}.
Here we limit attention to the dynamic quantities.

\subsection{Integrated autocorrelation times} \label{sec_data_analysis_tauint}

Let us begin by summarizing the qualitative behavior of
the integrated autocorrelation times $\tau_{{\rm int},{\cal O}}$
for different observables $\scro$,
as reported in Tables~\ref{table_dynamic_1} and \ref{table_dynamic_2}.
The three ``energy-like'' observables $\scrn,\scre,\scre'$ satisfy
\be
  \tau_{{\rm int},{\cal N}}  \;\le\;
  \tau_{{\rm int},{\cal E}}  \;\le\;
  \tau_{{\rm int},{\cal E}'}
 \label{ineq.energy-like}
\ee
and the two ``susceptibility-like'' observables $\scrm^2, \scrs_2$ satisfy
\be
  \tau_{{\rm int},{\cal M}^2}  \;\le\;
  \tau_{{\rm int},{\cal S}_2}
  \;,
 \label{ineq.sus-like}
\ee
in accordance with a rigorous theorem \cite{Li_Sokal,Salas_Sokal_Potts3}.
These five observables all have autocorrelation times
in the same ballpark, as do $\scrc_1$ and $\scrs_0$.
The autocorrelation times of $\scrc_2$ and $\scrc_3$,
by contrast, are notably smaller.
Of all the observables we measured,
$\scre'$ exhibits the largest autocorrelation time,
with $\scre$ and $\scrs_2$ only slightly behind.

Let us now fit the integrated autocorrelation times for
all these observables to a simple power law
$\tau_{{\rm int},{\cal O}} = A L^{z_{{\rm int},{\cal O}}}$.
We shall show the case of $\scre'$ in detail;
all the other observables behave similarly.
% {\bf Rewrite this section using the data of Table \ref{table_results}.
%    Start with $\scre'$ in detail, then do $\scre$ and $\scrn$ briefly;
%    then do $\scrs_2$ in detail and $\scrm^2$ briefly;
%    then do $\scrs_0$ and $\scrc_1$ briefly!!!!!}

\begin{figure}
\hspace*{-2cm}
\includegraphics[width=500pt]{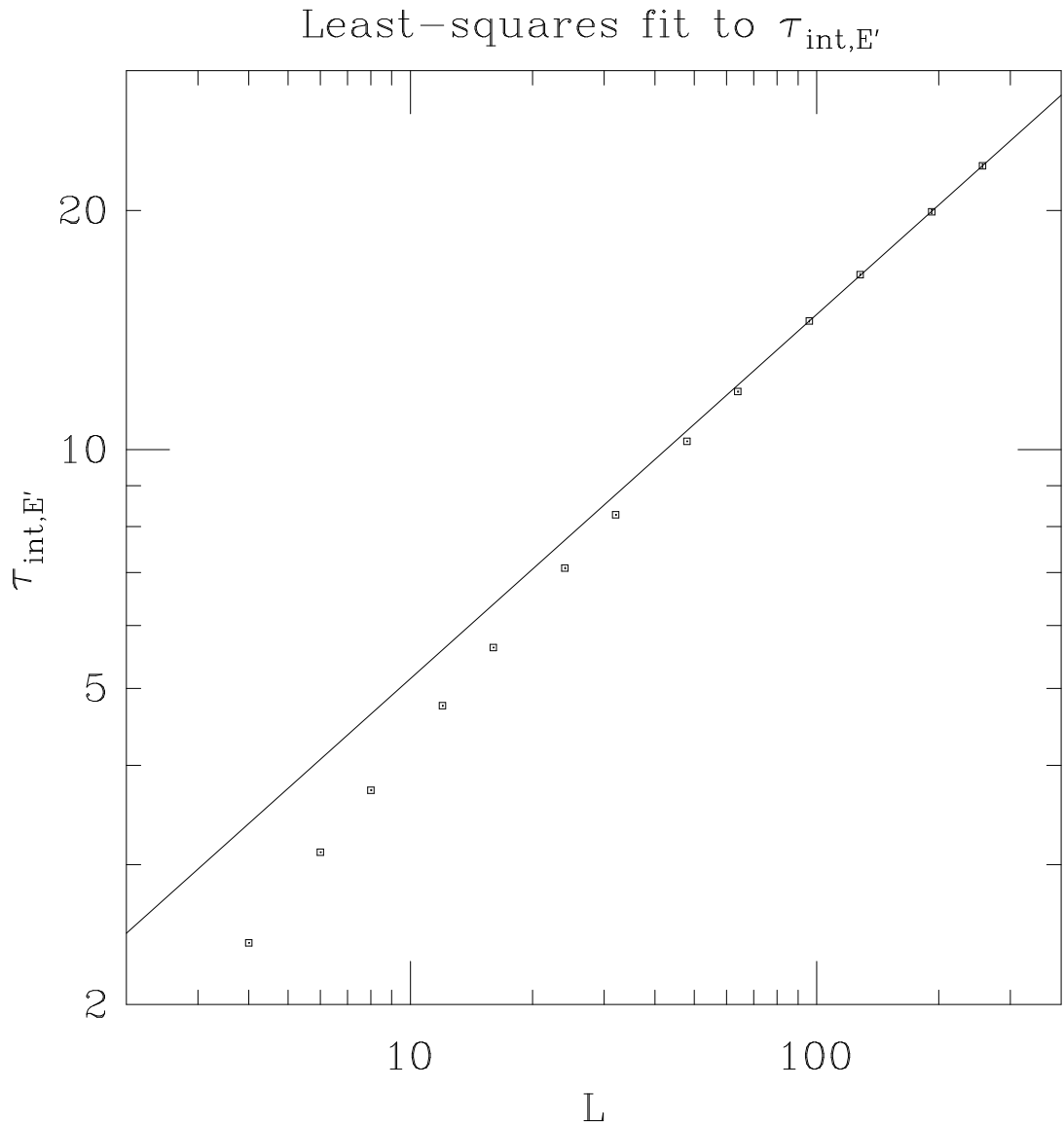}
\vspace*{-2cm}
\caption{
    $\tau_{{\rm int},{\cal E}'}$ versus $L$.
    Error bars are in all cases smaller than the plot symbol.
    The least-squares fit line is
    $\tau_{{\rm int},{\cal E}'} = 1.78861 L^{0.45878}$
    and is obtained for $L_{\rm min} = 96$.
}
\label{figure_taufit}
\end{figure}

In Figure~\ref{figure_taufit} we have made a log-log plot of
$\tau_{{\rm int},{\cal E}'}$ versus $L$.
(Please note that the error bars are significantly smaller
 than the plot symbols.)
The plot shows notable curvature,
i.e.\ there are fairly strong corrections to scaling,
at least for $L \ltapprox 64$.
Consequently, the least-squares fits with $L_{\rm min} \le 64$
all have enormous $\chi^2$ (confidence level $< 0.05\%$),
reflecting the fact that for $L \ltapprox 64$
the corrections to scaling are many times our (very small) error bars.
For $L_{\rm min} \ge 96$, by contrast, the $\chi^2$ values are good,
reflecting the fact that in this regime
the corrections to scaling are comparable to or smaller than our error bars.
Our preferred fit corresponds to $L_{\rm min} = 96$, and yields
\be
   z_{{\rm int},{\cal E}'} \;=\; 0.4588 \pm 0.0047
\label{z_int_Ep}
\ee
with $\chi^2=0.352$ (2 DF, level = 83.8\%);
here the error bar is one standard deviation
(i.e.\ confidence level $\approx 68\%$).

A similar pattern is obtained for all the other observables,
with the curvature always in the same direction.
In all cases our preferred fit corresponds to $L_{\rm min} = 96$;
the results of these fits are reported in Table~\ref{table_results}.
All the observables except $\scrc_2$ and $\scrc_3$
have exponents $z_{\rm int}$ in the vicinity $0.45 \pm 0.03$.
It is conceivable that the true values of these exponents
are in fact {\em exactly}\/ equal;
we do not know whether the small differences between the estimates
represent real differences or are merely the residual effects
of corrections to scaling.

It is worth noting that the rigorous inequality \reff{ineq.energy-like}
implies
\be
   z_{{\rm int},{\cal N}} \;\le\; z_{{\rm int},{\cal E}}
      \;\le\; z_{{\rm int},{\cal E}'}
   \;,
\ee
while the estimates in Table~\ref{table_results}
show the {\em opposite}\/ behavior.
This strongly suggests that in fact we have
\be
   z_{{\rm int},{\cal N}} \;=\; z_{{\rm int},{\cal E}}
      \;=\; z_{{\rm int},{\cal E}'}  \;,
\ee
in accordance with the ``almost-theorem'' proven in
\cite[Section 2.2]{Salas_Sokal_Potts3},
and that the deviations in Table~\ref{table_results}
result from corrections to scaling.
Unfortunately, we don't know which of these estimates
is closest to the true value;
but we are inclined to trust more the
estimate coming from the slowest of these modes, i.e.\ $\scre'$.
We therefore give as our final estimate
\be
   z_{{\rm int},{\cal N}} \;=\; z_{{\rm int},{\cal E}}
      \;=\; z_{{\rm int},{\cal E}'}  \;=\;
   0.459 \pm 0.005 \pm 0.025  \;,
 \label{estimate_energy-like}
\ee
where the first error bar represents statistical error
(68\% confidence interval)
and the second error bar
represents possible systematic error due to
the residual effects of corrections to scaling
(68\% subjective confidence interval).

The susceptibility-like observables $\scrm^2$ and $\scrs_2$,
by contrast, do show the correct inequality arising from \reff{ineq.sus-like}.
Comparing the two estimated exponents,
and bearing in mind the ``almost-theorem'' that they should be equal,
we give as our final estimate
\be
   z_{{\rm int},{\cal M}^2}  \;=\; z_{{\rm int},{\cal S}_2}  \;=\;
   0.443 \pm 0.005 \pm 0.030  \;.
 \label{estimate_sus-like}
\ee

The estimates \reff{estimate_energy-like} and \reff{estimate_sus-like}
are consistent with each other, as well as with the estimates for
$z_{{\rm int},{\cal S}_0}$ and $z_{{\rm int},{\cal C}_1}$.
This suggests that the true values of all these exponents might be
exactly equal.
Only the estimates for $z_{{\rm int},{\cal C}_2}$ and 
$z_{{\rm int},{\cal C}_3}$ are significantly lower than the others;
and even here, it is conceivable that the discrepancy again arises
from corrections to scaling.

% In conclusion, we have shown numerically
% that the three ``energy-like'' observables $\scrn,\scre,\scre'$
% all have the same dynamic critical exponent $z_{\rm int}$,
% as do the ``susceptibility-like'' observables $\scrm^2, \scrs_2$;
% this is in accordance with the ``almost-theorem'' proven in
% \cite[Section 2.2]{Salas_Sokal_Potts3}.
% It is not, however, clear whether these two dynamic critical exponents
% are exactly equal or not;
% it is possible that $z_{\rm int}$ for the
% ``susceptibility-like'' observables is very slightly smaller
% (by $\approx 0.02$).
% {\bf Well, they seem equal within the error given by corrections to scaling,
%    i.e.\ the discrepancies \emph{among} the energy-like and
%    susceptibility-like  observables.}

\subsection{Exponential autocorrelation time and autocorrelation functions}
 \label{sec_data_analysis_tauexp}

Recall that exponential autocorrelation time
of an observable ${\cal O}$ is defined as
\be
\tau_{{\rm exp},{\cal O}}  \;=\;
   \lim_{t \to \infty}  {-|t| \over  \log  \rho_{\cal OO}(t)}
   \;,
\ee
and that the exponential autocorrelation time of the system is defined as
\be
   \tau_{\rm exp}  \;=\;  \sup_{{\cal O}}  \tau_{{\rm exp},{\cal O}}  \;.
\ee
All observables that are not orthogonal to the system's slowest mode
satisfy $\tau_{{\rm exp},{\cal O}} = \tau_{\rm exp}$.
Since all the observables studied in this paper
are invariant under the symmetry group of the Potts model,
we have no reason to expect that any of them are orthogonal
to the slowest mode.
We therefore expect --- and will verify numerically ---
that they all have the {\em same}\/
exponential autocorrelation time $\tau_{{\rm exp},{\cal O}}$,
which is presumably equal to $\tau_{\rm exp}$.

We shall begin by discussing the qualitative behavior
of the autocorrelation functions for various observables at fixed $L$.
Then we shall discuss the $L$-dependence of various quantities
associated to the exponential decay of the autocorrelation functions.
Finally, in the next subsection, we shall discuss
the finite-size scaling of the autocorrelation functions.

\begin{figure}
\hspace*{-1.5cm}
\includegraphics[width=500pt]{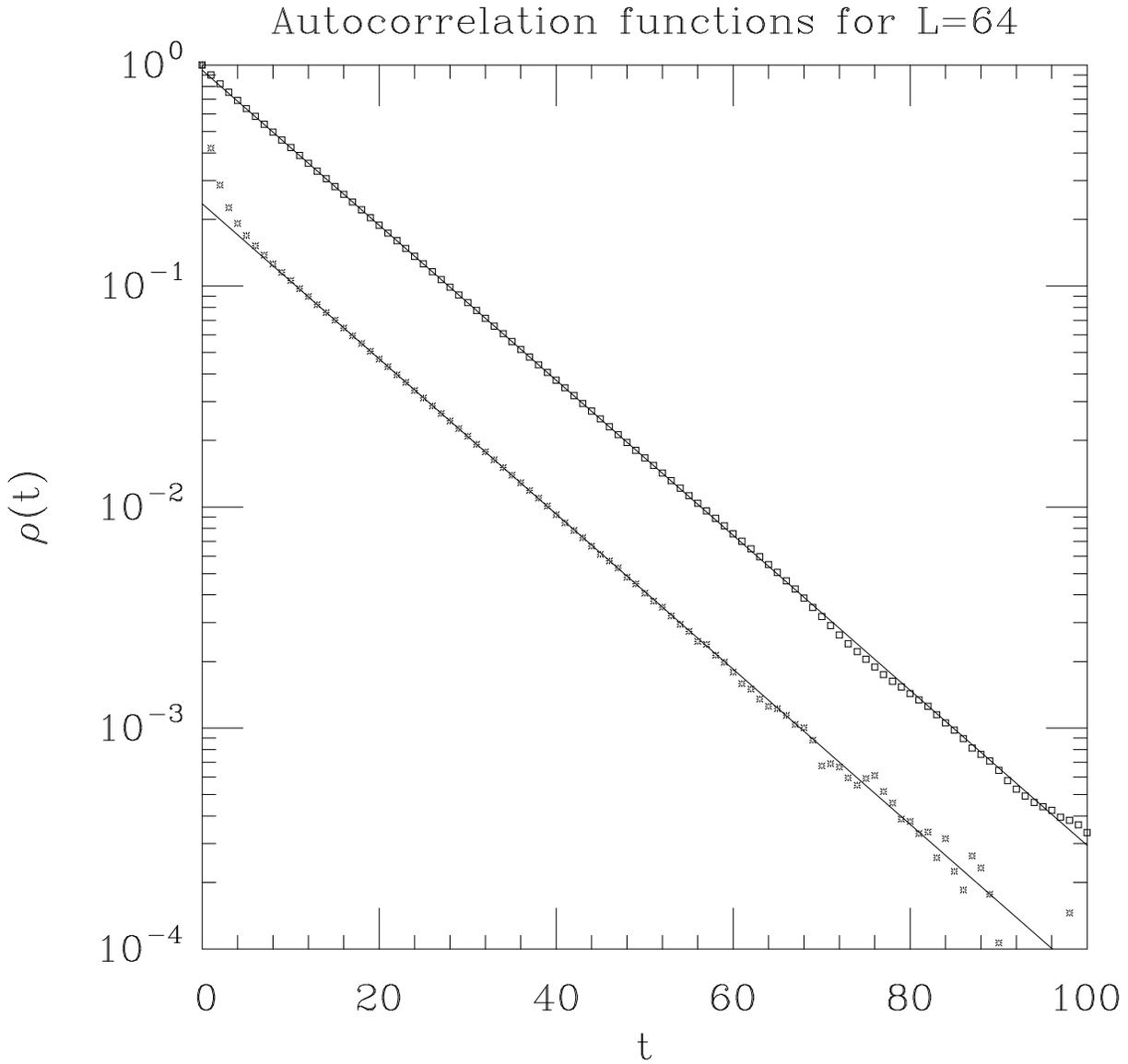}
\vspace*{-2cm}
\caption{
    Autocorrelation functions $\rho_{\cal OO}(t)$
    for observables ${\cal O} = {\cal E}'$ ($\Box$) and ${\cal C}_2$ ($\ast$)
    at $L=64$.
    We have also depicted the straight-line fits corresponding to
    $\tau_{{\rm exp},{\cal O}}$.
}
\label{figure_rhoEp}
\end{figure}

The typical behavior of the autocorrelation functions $\rho_{\cal OO}(t)$
is depicted in Figure~\ref{figure_rhoEp}.
For simplicity we have shown only the two observables
exhibiting the most extreme behavior (among these we have measured):
namely, $\scre'$, which has the largest $\tau_{{\rm int}}$
and whose autocorrelation function shows the least curvature
(i.e.\ is closest to a pure exponential);
and $\scrc_2$, which has the smallest $\tau_{{\rm int}}$
and whose autocorrelation function shows the most curvature.
The plots for all other observables are intermediate
between these two.\footnote{
   These behaviors underlie our choice of the window factors
   for estimating $\tau_{{\rm int},{\cal O}}$:  namely,
   $c=8$ for $\scrn$, $\scre$, $\scre'$, $\scrm^2$, $\scrs_2$ and $\scrc_1$,
   whose autocorrelation functions are close to a pure exponential;
   $c=10$ for $\scrs_0$; and $c=15$ for $\scrc_2$ and $\scrc_3$.
}
Clearly, each autocorrelation function behaves
asymptotically for large $t$ as
\be
   \rho_{\cal OO}(t)  \;\approx\;
   A_{\cal O} e^{-|t|/\tau_{{\rm exp},{\cal O}}}
   \;.
\ee
We obtained rough estimates of
$\tau_{{\rm exp},{\cal O}}$ and the amplitude $A_{\cal O}$
by performing an unweighted least-squares fit to
\be
   \log \rho_{\cal OO}(t)  \;=\;  a - bt
\ee
over the range
$\tau_{{\rm int},\scre'} \le t \le 3\tau_{{\rm int},\scre'}$
where all the autocorrelation functions are approximately
a pure exponential;
this yields $\tau_{{\rm exp},{\cal O}} = 1/b$
and $A_{\cal O} = e^a$.\footnote{
   In principle we should have done a properly weighted least-squares fit,
   taking account of the covariance matrix \reff{cov_rho}
   among the estimates $\widehat{\rho}_{\cal OO}(t)$.
   But this is rather complicated,
   and we did not think it was worth the effort.
   As a result of this laziness, we are unable to quote
   any reliable error bars on $\tau_{{\rm exp},{\cal O}}$ or $A$.
}
The results are shown in Tables~\ref{table_exponential_1}
and \ref{table_exponential_2}.
Clearly, all the observables studied here have the same value of
$\tau_{{\rm exp},{\cal O}}$, as expected theoretically.
We shall use $\tau_{{\rm exp},{\cal E}'}$ from Table~\ref{table_exponential_1}
as our best estimate of $\tau_{{\rm exp}}$.

\begin{figure}
\hspace*{-2cm}
\includegraphics[width=500pt]{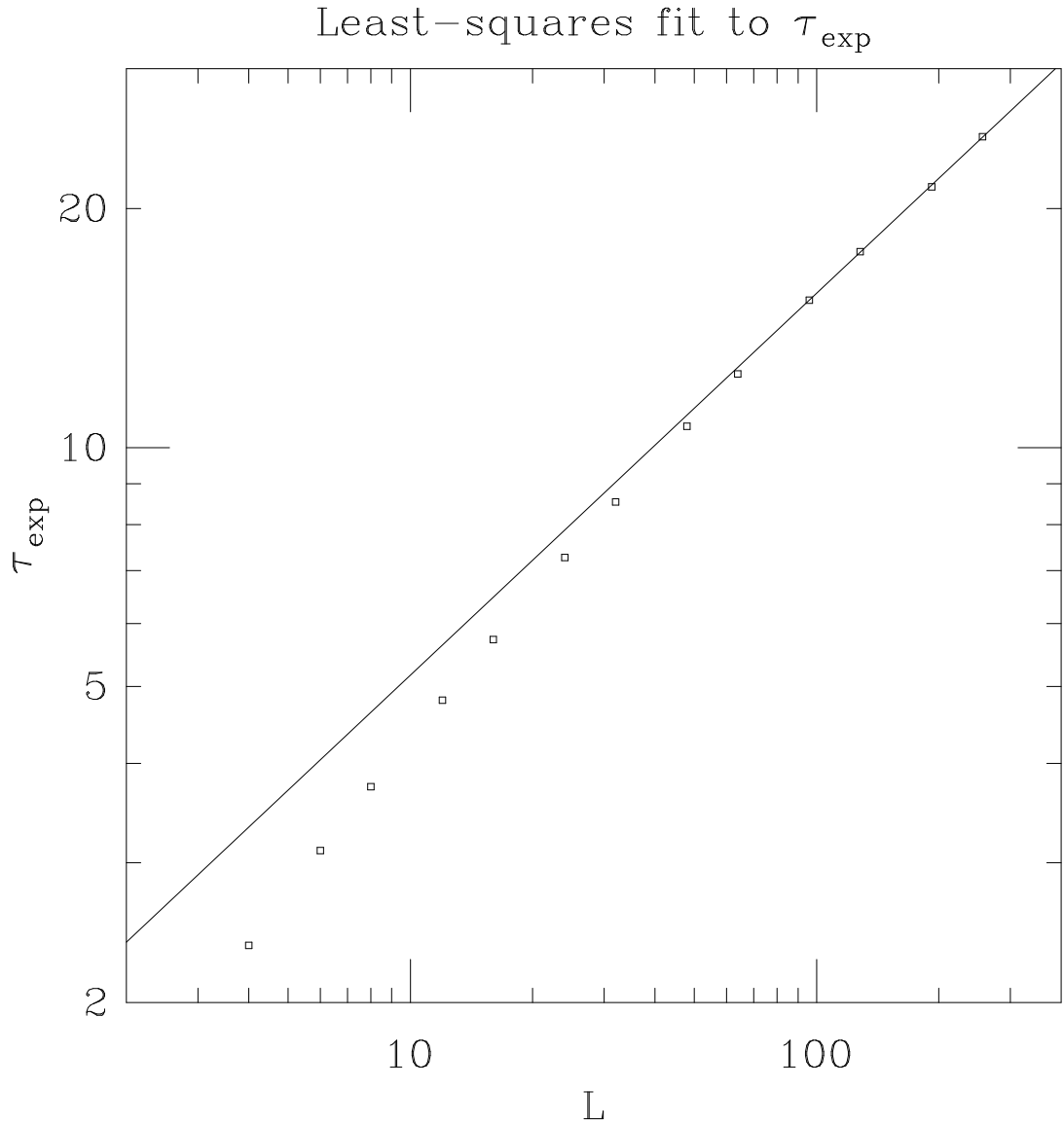}
\vspace*{-2cm}
\caption{
    $\tau_{{\rm exp}}$ versus $L$.
    The least-squares fit line is
    $\tau_{{\rm exp}} = 1.706 L^{0.481}$
    and is obtained for $L_{\rm min} = 96$.
    The value of $\tau_{{\rm exp}}$ is taken from
    $\tau_{{\rm exp},{\cal E}'}$ in Table~\ref{table_exponential_1}.
}
\label{figure_tauexpfit}
\end{figure}

In Figure~\ref{figure_tauexpfit} we have plotted
the estimated $\tau_{{\rm exp}}$ versus $L$.
We attempted to estimate the dynamic critical exponent $z_{\rm exp}$
by fitting
\be
   \tau_{{\rm exp}}  \;\approx\; B L^{z_{\rm exp}}
   \;.
\ee
In performing this fit, we used as rough error bars on $\tau_{{\rm exp}}$
the sum of the error bar on $\tau_{{\rm int},{\cal E}'}$
and the standard deviation of the $\tau_{{\rm exp}}$ estimates
for the seven observables $\scrn$, $\scre$, $\scre'$, $\scrm^2$, $\scrs_2$,
$\scrs_0$ and $\scrc_1$.
Our preferred fit has $L_{\rm min} = 96$,
and yields $z_{\rm exp} = 0.481 \pm 0.007$, $B = 1.706 \pm 0.058$
($\chi^2 = 0.754$, 2 DF, level = 68.6\%).
Of course, these error bars should not be taken terribly seriously.

It is not clear whether $z_{\rm exp}$ is equal to
$z_{{\rm int},{\cal E}'}$ or is slightly larger.
This question is related to the degree of curvature in
the plot of the autocorrelation function,
and more specifically to its $L$-dependence.
To investigate this question in more detail,
we first observed that
\be
   A_{\cal O} e^{-|t|/\tau_{\rm exp}}  \;\le\;
   \rho_{\cal OO}(t) \;\le\;
   e^{-|t|/\tau_{\rm exp}}
\ee
(as is obvious from Figure~\ref{figure_rhoEp}).
Therefore, if we define the modified autocorrelation time
\be
   \bar{\tau}_{\rm exp}
   \;\equiv\;
   {1 \over 2} \,
   \sum_{t=-\infty}^\infty e^{-|t|/\tau_{\rm exp}}
   \;=\;
   {1 \over 2} \,
   {1 + e^{-1/\tau_{\rm exp}}  \over  1 - e^{-1/\tau_{\rm exp}}}
   \;,
\ee
we necessarily have
\be
   A_{\cal O}
   \;\le\;
   {\tau_{{\rm int},{\cal O}}   \over   \bar{\tau}_{\rm exp} }
   \;\le\;
   1   \;.
 \label{AO_RO_ineq}
\ee
We therefore studied the $L$-dependence of the quantities
$A_{\cal O}$ and
$R_{\cal O} \equiv \tau_{{\rm int},{\cal O}}/\bar{\tau}_{\rm exp}$
for various observables ${\cal O}$
(see Tables~\ref{table_ratios_1} and \ref{table_ratios_2}).

We tried fits of $A_{\cal O}$ and $R_{\cal O}$
to the alternative Ans\"atze $c L^{-p}$ and
$c_1 + c_2 L^{-\omega}$ (with $\omega=0.82$).
Unfortunately, we do not have any valid error bars on
$A_{\cal O}$ and $R_{\cal O}$;
but we can assign fictitious error bars and compare the
{\em relative}\/ $\chi^2$ for the two fits.
We did this for the two extreme observables, $\scre'$ and $\scrc_2$.
In all cases we found that the power-law Ansatz
gives a much better fit, and also one that holds over a wider range of $L$.
We estimated $p \approx 0.023$ for $A_{\scre'}$,
$p \approx 0.021$ for $R_{\scre'}$,
$p \approx 0.092$ for $A_{\scrc_2}$,
$p \approx 0.135$ for $R_{\scrc_2}$.
The values for $p$ for $A_{\scre'}$ and $R_{\scre'}$ are very nearly equal,
and are almost exactly equal to our estimate of
$z_{\rm exp} - z_{{\rm int},{\cal E'}} \approx 0.022$.
As for the values for $p$ for $A_{\scrc_2}$ and $R_{\scrc_2}$,
they violate the rigorous inequality $p(A_{\scrc_2}) \ge p(R_{\scrc_2})$
that follows from \reff{AO_RO_ineq};
this strongly suggests that the two exponents
$p(A_{\scrc_2})$ and $p(R_{\scrc_2})$ are in fact equal,
though we do not know whether the correct value
lies nearer to 0.092 or to 0.135.
The latter value is fairly close to our estimate of
$z_{\rm exp} - z_{{\rm int},{\cal C}_2} \approx 0.127$.

Since for each of these observables it appears that
$A_{\cal O}$ and $R_{\cal O}$ have the same exponent $p$,
we tried fits of $R_{\cal O}/A_{\cal O}$ to the Ansatz
$c_1 + c_2 L^{-\omega}$ (with $\omega=0.82$).
For $\scre'$ we find a limiting value
$R_{\cal E'}/A_{\cal E'} \approx 1.010$;
for $\scrc_2$ it is more difficult to tell,
but a limiting value
$R_{{\cal C}_2}/A_{{\cal C}_2} \approx 1.14$ seems plausible.

\subsection{Finite-size scaling of autocorrelation functions}
 \label{sec_data_analysis_tauFSS}

A final way to study these questions is to investigate the
finite-size scaling of the autocorrelation functions.
The standard dynamic finite-size-scaling Ansatz
for the autocorrelation function $\rho_{\cal OO}(t)$ is
\be
\rho_{\cal OO}(t;L) \;\approx\; |t|^{-p_{\cal O}} h_{\cal O}\!\left(
          {t \over \tau_{{\rm exp},{\cal O}} }; {\xi(L) \over L} \right) \; .
\label{fss_rho}
\ee
(Here the dependence on the coupling constants,
 e.g.\ the inverse temperature,
 has been suppressed for notational simplicity.)
Summing \reff{fss_rho} over $t$, it follows that
\be
\tau_{{\rm int},{\cal O}} \;\sim\;
 \tau_{{\rm exp},{\cal O}}^{1 - p_{\cal O}} \; ,
\ee
or equivalently,
\be
z_{{\rm int},{\cal O}}  \;=\;  (1 - p_{\cal O}) z_{{\rm exp},{\cal O}} \; .
\ee
Thus, only when $p_{\cal O}=0$ do we have
$z_{{\rm int},{\cal O}} = z_{{\rm exp},{\cal O}}$
\cite{Sokal_Cargese}.
In this latter case the Ansatz \reff{fss_rho} can be rewritten in
the equivalent form
\be
\rho_{\cal OO}(t;L) \;\approx\; \widehat{h}_{\cal O}\!\left(
          {t \over \tau_{{\rm int},{\cal O}} }; {\xi(L) \over L} \right) \; .
\label{fss_rho_p=0}
\ee
%

\begin{figure}
\hspace*{-1.3cm}
\includegraphics[width=500pt]{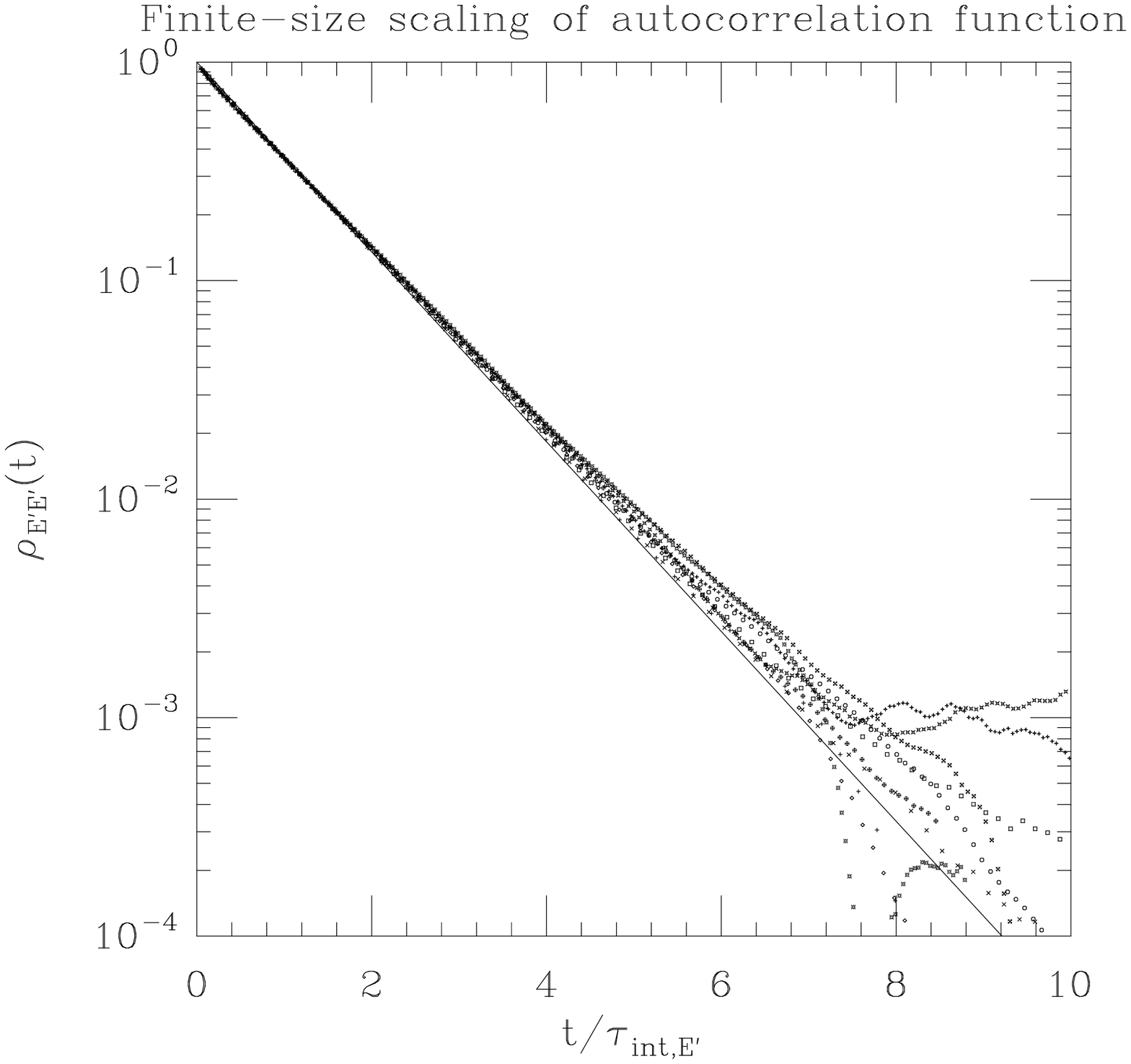}
\vspace*{-2cm}
\caption{
    $\rho_{\cal E'E'}(t)$ versus $t/\tau_{{\rm int},{\cal E'}}$ for
    $12 \leq L \leq 256$. The different symbols denote the different
    lattice sizes: $L=12$ ($+$), $L=16$ ($\times$), $L=24$ ($\Box$),
    $L=32$ ($\Diamond$), $L=48$ ($\circ$), $L=64$ ($\protect\fancyplus$),
    $L=96$ ($\protect\fancycross$), $L=128$ ($\protect\fancysquare$),
    $L=192$ ($\protect\fancydiamond$), $L=256$ ($\ast$).
    For comparison, we have also depicted the line corresponding to the pure
    exponential $\rho_{\cal E'E'}(t) = \exp(-t/\tau_{{\rm int},{\cal E'}})$.
}
\label{figure_fss_rhoEp}
\end{figure}

% {\bf The following is taken from my 3-state paper with Salas,
%    and needs to be adapted to this one.
%    ALAN: I have to try plots with $p_\scro \neq 0$!!!}
To test this latter Ansatz,
we have plotted $\log \rho_{\cal OO}(t)$ versus
$t/\tau_{{\rm int},{\cal O}}$ for
the observable ${\cal O} = {\cal E}'$ (Figure~\ref{figure_fss_rhoEp}).
For clarity we have included only the data from $L \ge 12$;
the data coming from different lattice sizes are plotted
with different symbols.
We have also depicted for reference
a line corresponding to the pure exponential
$\rho_{\cal E'E'}(t) = e^{-t/\tau_{{\rm int},{\cal E'}}}$.

The data fall roughly onto a single curve,
but there are clear corrections to scaling:
the points move upwards (away from the pure exponential line)
as $L$ increases.
It is not clear whether the points are tending to a limiting curve
as $L \to\infty$, or whether they will continue indefinitely to move upwards.
This is another way of saying that we do not know whether
$p_{{\cal E'}} = 1 - z_{{\rm int},{\cal E}'}/z_{\rm exp}$
is exactly zero or is slightly positive (e.g. $\approx 0.067$).

\section{Discussion}  \label{sec_discussion}

In this paper we have obtained high-precision data
at the critical point of the three-dimensional Ising model
on fairly large lattices (up to $L=256$),
which have allowed us to derive quite accurate estimates of the
dynamic critical exponents $z_{{\rm int},{\cal O}}$ and $z_{\rm exp}$
for the Swendsen--Wang algorithm for this model.
Our data resolve the discrepancies between previous works
\cite{Swendsen_87,Wolff_89,Wang_90,Coddington_92,%
Kerler_93a,Kerler_93b,Hennecke_93,Wang_02},
which can now be understood as arising from corrections to scaling.

We would like to conclude by comparing our numerical results with
some of the theoretical frameworks that have been proposed by previous authors.
These frameworks have as their goal
to understand the dynamic critical behavior of the
Swendsen--Wang algorithm for the various ferromagnetic Potts models,
and in particular to relate the dynamic critical exponent(s) $z_{\rm SW}$
to the {\em static}\/ critical exponents for the same models.
The three most important of these frameworks are
the Li--Sokal proof \cite{Li_Sokal}
and its extensions \cite{Salas_Sokal_Potts3,Salas_Sokal_AT};
the scaling Ansatz of Klein, Ray and Tamayo \cite{Klein_89};
and the empirically based conjectures of
Coddington and Baillie \cite{Coddington_92}.

We have discussed the Li--Sokal bound $z_{\rm SW} \ge \alpha/\nu$
in the Introduction, and there is not much more to say.
Suffice it to observe once again that while this bound is close to sharp
(and possibly even sharp modulo a logarithm)
for the two-dimensional Potts models with $q=2,3,4$,
it is clearly far from sharp in the three- and four-dimensional
Ising models.
Our data for the three-dimensional Ising model
yield $z_{\rm SW} \approx 0.46$,
compared to $\alpha/\nu \approx 0.1756$ \cite{Hasenbusch_99};
and for the four-dimensional Ising model
it is generally believed that $z_{\rm SW}=1$
\cite{Klein_89,Ray_89,Coddington_92,Persky_96,Jerrum_in_prep},
compared to $\alpha/\nu = 0$ $(\times \log^{1/3})$.
Clearly, some other physical mechanism,
beyond the one exploited in the Li--Sokal proof,
must be principally responsible for the critical slowing-down
in these latter models;
the central open problem is to identify this mechanism
and to determine theoretically the dynamic critical exponent.

Klein, Ray and Tamayo \cite{Klein_89} have presented a scaling Ansatz
leading to the conjecture
\be
   z_{\rm SW}  \;=\;  z_{\rm G}  \,-\, {2 \gamma \over d_m \nu}
   \;,
 \label{eq.KRT.1}
\ee
where $z_{\rm G}$ is the dynamic critical exponent for the
Glauber dynamics in the same model,
and $d_m$ is ``the mean fractal dimension of the finite clusters''
in the Fortuin--Kasteleyn representation of the model.\footnote{
   The discussion of Klein, Ray and Tamayo \cite{Klein_89}
   does not distinguish between the various dynamic critical exponents
   $z_{{\rm int},{\cal O}}$ and $z_{\rm exp}$.
   For simplicity we can assume that their discussion refers to $z_{\rm exp}$
   for both the Swendsen--Wang and Glauber algorithms.
}
The trouble with this Ansatz, alas, is not simply that the
numerical value of $d_m$ is unknown;
it is, rather, that the definition of $d_m$ is too vague
to serve even as a guide for numerical attempts to determine its value.
(The situation would be different if, for example, $d_m$ could be defined
 as a dimension associated with the scaling behavior of some specific
 observable.)
Consequently, Klein, Ray and Tamayo were limited in practice to
observing that $d_m$ presumably ``lies between $d$, the spatial dimension,
and $d_f = d - \beta/\nu$, the fractal dimension of the incipient
infinite cluster'' \cite[p.~164]{Klein_89}.
This plausible reasoning yields the conjectured {\em inequality}\/
\be
   z_{\rm G}  \,-\, {2 \gamma \over d \nu - \beta}
   \;\le\;
   z_{\rm SW}
   \;\le\;
   z_{\rm G}  \,-\, {2 \gamma \over d \nu}
   \;.
 \label{eq.KRT.2}
\ee
In Table~\ref{table_discussion} we compare $z_{\rm SW}$
with the Klein--Ray--Tamayo bounds
$z_{\rm KRT}^{\rm (lower)}$ and $z_{\rm KRT}^{\rm (upper)}$
for the ferromagnetic Potts models (in dimension $d \le 4$)
having a second-order transition.
The bounds appear to be violated for all three two-dimensional Potts models:
for the Ising and 3-state models, the violation is possibly within the
errors in the determination of $z_{\rm SW}$ and $z_{\rm G}$,
especially if one takes into account the possibility of logarithms;
but for the 4-state model, the violation is blatant
(barring a gross error in the determination of $z_{\rm G}$).
For the three-dimensional Ising model, curiously,
the lower bound seems to be exact (within errors);
while for the four-dimensional Ising model,
the upper bound is exact (modulo logarithms).
It would be interesting to know whether the latter facts
are anything more than curious coincidences.

Coddington and Baillie \cite{Coddington_92}
carried out a careful numerical study of the
dynamic critical behavior of the Swendsen--Wang algorithm
for the Ising models in dimensions $d=2,3,4$,
on the basis of which they made the remarkable conjecture that
for these models $z_{\rm SW} = \beta/\nu$ {\em exactly}\/.
More specifically, they observed that the mean size of the largest cluster,
here denoted $C_1$, scales at the critical point as
$C_1 \sim L^{d - \beta/\nu}$,
and they found that their data could be explained by the asymptotic Ansatz
\be
   \tau_{\rm SW}  C_1 / L^d  \;\approx\;  a + b \log L
   \;.
\ee
If true, this would imply that the correct dynamic critical exponent
$z_{\rm SW}$ for the two-dimensional Ising model
is neither 0 $(\log)$ \cite{Heermann_90} nor
$0.222 \pm 0.007$ \cite{Salas_Sokal_Ising_v1},
but rather 1/8 (possibly multiplied by a logarithm).
The data of \cite{Salas_Sokal_Ising_v1} are certainly consistent
with this possibility, though they do not distinguish it from the
other possible behaviors.\footnote{
   Unfortunately, $C_1$ was not measured in \cite{Salas_Sokal_Ising_v1}.
}

\begin{figure}
\hspace*{-2cm}
\includegraphics[width=500pt]{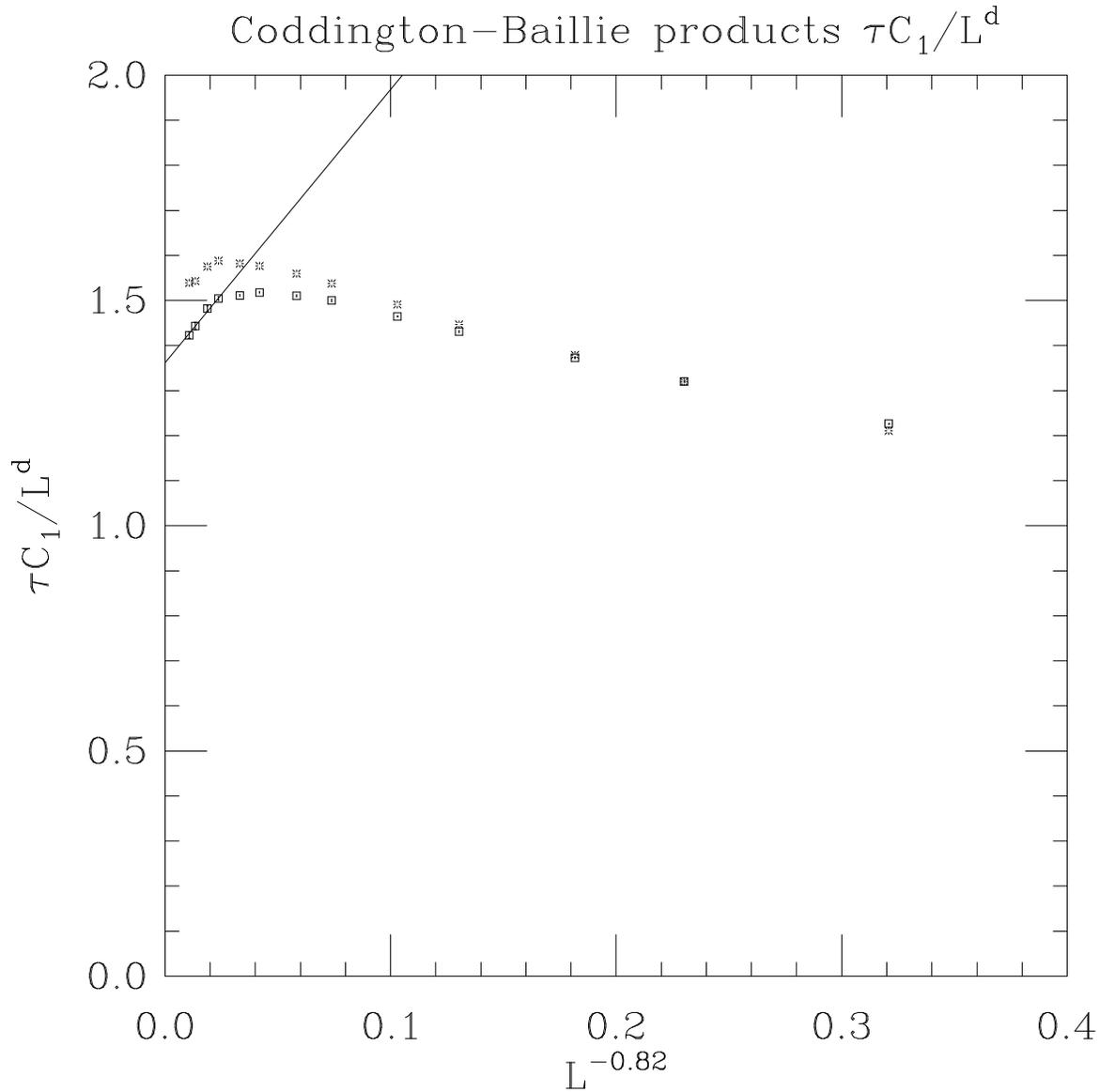}
\vspace*{-2cm}
\caption{
    The Coddington--Baillie products
    $\tau_{{\rm int},{\cal E}'} C_1 / L^d$
    ($\Box$ with triangle-inequality error bars)
    and $\tau_{{\rm exp},{\cal E}'} C_1 / L^d$
    ($\ast$ without error bars)
    versus $L^{-0.82}$.
    The least-squares fit line is
    $\tau_{{\rm int},{\cal E}'}  C_1 / L^d = 1.362 + 6.064 L^{-0.82}$
    and is obtained for $L_{\rm min} = 96$.
}
\label{figure_coddington}
\end{figure}

For the three-dimensional Ising model,
the Coddington--Baillie conjecture would imply that $z_{\rm SW} = 0.5183(4)$,
which at first sight is incompatible at the $3\sigma$ level
with our estimate $z_{{\rm int},{\cal E}'} = 0.459 \pm 0.005 \pm 0.025$
(central value $\pm$ statistical error $\pm$ systematic error).
Indeed, the curvature in Figure~\ref{figure_taufit},
assuming that it continues in the same direction,
suggests that the true $z_{{\rm int},{\cal E}'}$ is, if anything,
slightly {\em lower}\/ than our estimate based on $L \le 256$.
On the other hand, the difference between these exponents is small,
and a small exponent is very difficult to distinguish from zero.
We therefore made a direct test of the Coddington--Baillie conjecture
by studying the combination $\tau_{{\rm int},{\cal E}'}  C_1 / L^d$.
(Since we don't have statistically valid error bars for this combination,
 we used the triangle inequality to set worst-case error bars.)
A fit to $\tau_{{\rm int},{\cal E}'} C_1 / L^d = AL^p$
yields a decent $\chi^2$ only for $L_{\rm min} \ge 96$;
our preferred fit is $L_{\rm min} = 96$
and yields $p = -0.0573 \pm 0.0052$,
$\log A = 0.670 \pm 0.025$
($\chi^2 = 0.261$, 2 DF, level = 87.8\%).
This estimate for $p$ is, not surprisingly, in almost perfect agreement
with the values $z_{\rm SW} = 0.459$ and $\beta/\nu = 0.5183$.\footnote{
   Just to be safe,
   we also checked the theoretically predicted scaling of $C_1$.
   Using the Ansatz $C_1/L^d = A L^p$,
   we get a good fit already with $L_{\rm min} = 48$:
   $p = -0.5162 \pm 0.0002$,
   $\log A = 0.0892 \pm 0.0008$
   ($\chi^2 = 5.848$, 4 DF, level = 21.1\%).
   This value is not, strictly speaking, consistent with the
   estimate $\beta/\nu = 0.5183(4)$,
   but it is very close;
   the difference 0.002 can surely be understood as an effect
   of the residual corrections to scaling.
}
On the other hand, if we fit to
$\tau_{{\rm int},{\cal E}'} C_1 / L^d = A + B L^{-\omega}$
with $\omega = 0.82$,
a decent $\chi^2$ is again obtained only for $L_{\rm min} \ge 96$;
our preferred fit is again $L_{\rm min} = 96$
and we get $A = 1.362 \pm 0.011$,
$B = 6.064 \pm 0.553$
($\chi^2 = 1.726$, 2 DF, level = 42.2\%).
Figure~\ref{figure_coddington} shows
the data points ($\Box$) and the corresponding fit.
The fact that a reasonable fit is obtained with $A$ far from zero
(on the scale set by the observed values of
 $\tau_{{\rm int},{\cal E}'} C_1 / L^d$)
means that our data are also consistent with
a behavior $\tau_{{\rm int},{\cal E}'} C_1/L^d \to A > 0$ as $L \to\infty$,
and hence $p=0$.

It is very hard (if not impossible) to distinguish,
on purely numerical grounds, between these two behaviors.
The $\chi^2$ is slightly better for the power-law fit,
but this minor difference should not be taken terribly seriously.
The bottom line, it seems to us, is this:
the data shown in Figure~\ref{figure_coddington}
do not give any strong reason to believe that
$\tau_{{\rm int},{\cal E}'} C_1 / L^d$
is tending to {\em zero}\/ as $L \to\infty$.
Indeed, inspection of the curve would suggest a limit in the range 1.2--1.3,
depending on the extent to which the curvature continues at larger $L$;
this predicted limit is only about 20\% below
the maximum value attained at $L \approx 40$, and is thus very far from zero.
A more reliable judgment on the limiting value of
$\tau_{{\rm int},{\cal E}'} C_1 / L^d$
will have to wait 5--10 years,
when data will hopefully be available at (say) $L=512$ and $L=1024$.
But as things stand today, our data are fully consistent with the
Coddington--Baillie conjecture (albeit without a logarithmic term).

If the Coddington--Baillie conjecture is interpreted as applying
to $z_{\rm exp}$ rather than to $z_{{\rm int},{\cal E}'}$,
then the consistency between our data and the conjecture is even stronger.
This is to be expected, as our estimate $z_{\rm exp} \approx 0.481$
is closer to the value $\beta/\nu \approx 0.5183$.
The data points for $\tau_{{\rm exp},{\cal E}'} C_1 / L^d$
are also shown (alas, without error bars)
in Figure~\ref{figure_coddington} (points $\ast$).
The curvature is slightly weaker than for
$\tau_{{\rm int},{\cal E}'} C_1 / L^d$,
and the data seem to be tending to a limit in the range 1.35--1.45.

Of course, as Coddington and Baillie \cite{Coddington_92} themselves observe,
$z_{\rm SW} = \beta/\nu$ cannot possibly be a general identity
for the Swendsen--Wang dynamics,
as it clearly fails for the 2-dimensional Potts models
with $q=3$ and $q=4$ (see Table~\ref{table_discussion}).
Indeed, the Li--Sokal bound \reff{Li_Sokal_bound} ensures that
we must have $z_{\rm SW} > \beta/\nu$ in any Potts model where
$\alpha/\nu > \beta/\nu$.
At best, the identity $z_{\rm SW} = \beta/\nu$ could hold
for the special case of the Ising models.

Since the Swendsen--Wang algorithm is defined naturally for
all ferromagnetic Potts models,
a theoretical framework that is valid only for the Ising case
seems unnatural and, in our opinion, unlikely to be correct.
But the Coddington--Baillie conjecture can be rephrased in the
following way so as to be potentially valid for all Potts ferromagnets.
Suppose that there exists an as-yet-not-understood physical mechanism
causing slowness of the Swendsen--Wang dynamics
that is somehow related to the typical size of the largest cluster.
In this case, an inequality of the form
\be
  \tau_{\rm SW}  \;\ge\; {\rm const} \times {L^d \over C_1}
   \;,
 \label{ineq.CB}
\ee
analogous to the Li--Sokal bound \reff{Li_Sokal_bound_CH},
might hold for all Potts ferromagnets,
irrespective of dimension and number of states.
Indeed, it might even be possible to prove such an inequality rigorously
(for one or another of the various autocorrelation times),
if the physical basis were sufficiently well understood.
Furthermore, it is even conceivable that the Li--Sokal mechanism
and this new mechanism might together {\em exhaust}\/
the reasons for slowness in the Swendsen--Wang dynamics,
leading to the exact relation
\be
   z_{\rm SW}  \;=\;  \max(\alpha/\nu, \beta/\nu)
 \label{eq.grand_conj}
\ee
(possibly modulo a logarithm) for all Potts ferromagnets.
All currently available numerical data are consistent with
the validity of the grand conjecture \reff{eq.grand_conj},
provided that a multiplicative logarithm is permitted but not mandatory.

This conjecture is, of course, a wild speculation;
indeed, we consider it unlikely, {\em a priori}\/,
for a dynamic critical exponent of any nontrivial dynamics
to be exactly expressible in terms of static critical exponents
(except for trivial cases such as Gaussian models).
But stranger things have happened;
and this conjecture is, in any case, certainly worth closer investigation.
More modestly, this line of reasoning suggests that
efforts be made to prove the inequality \reff{ineq.CB}
and to understand what kind of physical mechanism might cause it to hold.

%
%END TEXT SECTIONS
%
\section*{Acknowledgments}

We wish to thank Henk Bl\"ote, Lu\'{\i}s Antonio Fern\'andez,
Werner Kerler, Lev Shchur and Jian-Sheng Wang
for valuable correspondence and for sharing their unpublished data;
Mulin Ding and Raj Sivanandarajah for efficient assistance
with numerous aspects of the Physics Department computing system;
and Rom\'an Scoccimarro for generously letting us
do some test runs on his DEC Alpha computer.

This research was supported in part by
U.S.\ National Science Foundation grants PHY--0099393
and PHY--0116590. %% This is the MRI computer grant

%%%\newpage
%
\renewcommand{\baselinestretch}{1}
\large\normalsize
%
%
%
%%%%%%%%%%%%   references  %%%%%%%%%%%%%%%%%%%%%%%%
%

\clearpage

%%%%%%%%%%%% START OF TABLES %%%%%%%%%%%%%
\newpage
\def\kk{\phantom{1}}

%
% TABLE 1: SUMMARY OF RUNS
%

%
% summary of the runs
%\label{sum_runs}
%
\begin{table}%[hb]
%\centering
%\vspace*{-0.5in} % Move table upwards
\addtolength{\tabcolsep}{-1.0mm}
\hspace*{-4mm}    % Move table leftwards
\protect\small
\begin{tabular}{|r|r|r|c|r|c|c|c|}
\hline
  \multicolumn{1}{|c|}{$L$}  &
  \multicolumn{1}{|c|}{Total \#}  &
  \multicolumn{1}{|c|}{Total \#}  &
  \multicolumn{1}{|c|}{Lengths (and numbers)}  &
  \multicolumn{1}{|c|}{$\tau_{{\rm int},{\cal E'}}$}  &
  \multicolumn{1}{|c|}{\# measurements}  &
  \multicolumn{1}{|c|}{CPU time}  &
  \multicolumn{1}{|c|}{CPU time}
  \\
  \multicolumn{1}{|c|}{\quad}  &
  \multicolumn{1}{|c|}{iterations}  &
  \multicolumn{1}{|c|}{discarded}  &
  \multicolumn{1}{|c|}{of individual runs}  &
  \multicolumn{1}{|c|}{\quad}  &
  \multicolumn{1}{|c|}{in units of $\tau_{{\rm int},{\cal E'}}$}  &
  \multicolumn{1}{|c|}{($\mu$sec/spin)}  &
  \multicolumn{1}{|c|}{(years)}
  \\
\hline
%%% & & & & & &   \\
4   & $10^8$ &	$10^5$ & $10^8$ (1) &
            2.3873  & $4.2 \!\times\! 10^7$ &  0.90 &  \kk0.000 \\
6   & $10^8$ &  $10^5$ & $10^8$ (1) &
            3.1054  & $3.2 \!\times\! 10^7$ &  0.56 &  \kk0.000 \\
8   & $10^8$ &  $10^5$ & $10^8$ (1) &
            3.7182  & $2.7 \!\times\! 10^7$ &  0.45 &  \kk0.001 \\
12  & $10^8$ &  $10^5$ & $10^8$ (1) &
            4.7469  & $2.1 \!\times\! 10^7$ &  0.40 &  \kk0.002 \\
16  & $5 \!\times\! 10^8$ &  $5 \!\times\! 10^5$ & $10^8$ (5) &
            5.6180 & $8.9 \!\times\! 10^7$ & 0.39 & \kk0.025 \\
24  & $10^8$   &  $10^5$ & $10^8$ (1) &
            7.0712 & $1.4 \!\times\! 10^7$ &   0.40 & \kk0.018 \\
32  & $2\!\times\! 10^8$ & $2 \!\times\! 10^5$ & $10^8$ (2) &
            8.2563 & $2.4 \!\times\! 10^7$ &   0.42 & \kk0.087 \\
48  & $3.8 \!\times\! 10^8$ & $2.1 \!\times\! 10^6$ & $10^8$ (2), $10^7$ (17), &
           10.2101 & $3.7 \!\times\! 10^7$ & 0.43 & \kk0.573 \\
    &   &   & $5 \!\times\! 10^6$ (2) &  &  &  &  \\
64  & $3.1 \!\times\! 10^8$ & $2.7 \!\times\! 10^6$ & $2 \!\times\! 10^7$ (12), $10^7$ (5),&
           11.7976 & $2.6 \!\times\! 10^7$ & 0.44 & \kk1.133 \\
    &   &   & $2 \!\times\! 10^6$ (10) &  &  &  &  \\
96  & $8.6 \!\times\! 10^7$  & $1.9 \!\times\! 10^6$ &
 $5 \!\times\! 10^6$ (17), $5 \!\times\! 10^5$ (2) &
           14.4677 & $5.8 \!\times\! 10^6$ & 0.45 & \kk1.085 \\
128 & $4.8 \!\times\! 10^7$  & $2.4 \!\times\! 10^6$ & $2.5 \!\times\! 10^6$ (16), $10^6$ (8) &
           16.5379 & $2.8 \!\times\! 10^6$ & 0.46 & \kk1.467 \\
192 & $2.9 \!\times\! 10^7$  & $3.3 \!\times\! 10^6$ & $10^6$ (25), $5 \!\times\! 10^5$ (8)&
           19.8511 & $1.3 \!\times\! 10^6$ & 0.47 & \kk3.057 \\
256 & $4.05 \!\times\! 10^7$  & $4.6 \!\times\! 10^6$ & $10^6$ (35), $5 \!\times\! 10^5$ (11)&
           22.6743 & $1.6 \!\times\! 10^6$ & 0.48 & 10.335 \\
%%% & & & & & &   \\
\hline
\end{tabular}
\caption{
   Summary of our runs.
   The total CPU time used in these runs was approximately 17.8 years.
}
\label{sum_runs}
\end{table}

%% TABLE 2: STATIC QUANTITIES
%%
\begin{table}%[th]
%\centering
%\vspace*{-0.5in} % Move table upwards
\addtolength{\tabcolsep}{-1.0mm}
\hspace*{-1.2cm}    % Move table leftwards
\protect\small
\begin{tabular}{|r|r@{$\,\pm\,$}r|r@{$\,\pm\,$}r|%
                     r@{$\,\pm\,$}r|r@{$\,\pm\,$}r|r@{$\,\pm\,$}r|}%
\hline  \\[-5mm]
\multicolumn{1}{|c|}{$L$}    &
%\multicolumn{1}{c|}{$MCS$}  &
\multicolumn{2}{c|}{$\chi$}  &
\multicolumn{2}{c|}{$C_H$}   &
\multicolumn{2}{c|}{$\xi$}   &
\multicolumn{2}{c|}{$E$}     &
\multicolumn{2}{c|}{$C_1/L^d$}   \\[0.5mm]
\hline\hline
%    4 & {\rm exact}
%      &\multicolumn{2}{l|}{\kk\kk\kk\kk12.204711}
%      &\multicolumn{2}{l|}{\kk1.496719}
%      &\multicolumn{2}{l|}{\kk\kk3.862380}
%      &\multicolumn{2}{l|}{1.710062}  \\
%\hline
4   &   21.1944   &   0.0033   &   6.6297   &   0.0012   &   2.52000   &   0.00036   &   0.4310438   &   0.0000405   & 0.512987 & 0.000043   \\
6   &   49.0575   &   0.0089   &   8.3079   &   0.0019   &   3.81458   &   0.00057   &   0.3881242   &   0.0000280   & 0.424582 & 0.000043   \\
8   &   87.8535   &   0.0175   &   9.4539   &   0.0024   &   5.10153   &   0.00080   &   0.3690531   &   0.0000212   & 0.368769 & 0.000042   \\
12   &   197.8047   &   0.0449   &   11.0562   &   0.0032   &   7.66909   &   0.00130   &   0.3522118   &   0.0000140  & 0.300979 & 0.000039  \\
16   &   350.5792   &   0.0388   &   12.2219   &   0.0017   &   10.23802   &   0.00083   &   0.3448934   &   0.0000047  & 0.260170 & 0.000017 \\
24   &   783.1500   &   0.2170   &   13.9336   &   0.0050   &   15.38221   &   0.00302   &   0.3385100   &   0.0000068   & 0.211590 & 0.000034 \\
32   &   1382.5072   &   0.2921   &   15.1928   &   0.0042   &   20.52010   &   0.00303   &   0.3357384   &   0.0000035  & 0.182575 & 0.000023 \\
48   &   3076.1461   &   0.5232   &   17.0744   &   0.0038   &   30.80276   &   0.00359   &   0.3333254   &   0.0000016  & 0.148217 & 0.000015 \\
64   &   5420.9953   &   1.0953   &   18.4697   &   0.0049   &   41.08335   &   0.00563   &   0.3322826   &   0.0000013  & 0.127789 & 0.000015 \\
96   &   12038.7855   &   5.1149   &   20.5499   &   0.0114   &   61.65067   &   0.01750   &   0.3313744   &   0.0000016 & 0.103645 & 0.000026  \\
128   &   21193.9763   &   13.0416   &   22.1551   &   0.0179   &   82.20142   &   0.03354   &   0.3309822   &   0.0000016  & 0.089301 & 0.000033 \\
192   &   47036.4986   &   41.9075   &   24.4737   &   0.0288   &   123.33211   &   0.07213   &   0.3306421   &   0.0000013 & 0.072421 & 0.000039  \\
256   &   83001.9797   &   66.8752   &   26.2767   &   0.0280   &   164.74988   &   0.08662   &   0.3304971   &   0.0000008 & 0.062502 & 0.000031  \\
\hline
\end{tabular}
\caption{
   Static data from the Monte Carlo simulations at the critical point
   of the 3-dimensional Ising model.
   For each lattice size ($L$),
   we report the susceptibility ($\chi$),
   the specific heat ($C_H$),
   the second-moment correlation length ($\xi$),
   the energy ($E$),
   and the mean size of the largest cluster ($C_1$).
   The quoted error bar corresponds to one standard deviation
   (i.e.\ confidence level $\approx$ 68\%).
}
\label{table_static_full}
\end{table}

%%
%% TABLE 3: DYNAMIC QUANTITIES [part.I]
%%
\begin{table}%[h]
%\centering
%\vspace*{-0.5in} % Move table upwards
\addtolength{\tabcolsep}{-1.0mm}
\hspace*{-3mm}    % Move table leftwards
%\protect\footnotesize
\begin{tabular}{|r|r@{$\,\pm\,$}r|r@{$\,\pm\,$}r|%
                   r@{$\,\pm\,$}r|r@{$\,\pm\,$}r|r@{$\,\pm\,$}r|}%
\hline  \\[-0.5cm]
\multicolumn{1}{|c|}{$L$}    &
\multicolumn{2}{c|}{$\tau_{{\rm int},{\cal N}}$} &
\multicolumn{2}{c|}{$\tau_{{\rm int},{\cal E}}$}   &
\multicolumn{2}{c|}{$\tau_{{\rm int},{\cal E}^\prime}$}   &
\multicolumn{2}{c|}{$\tau_{{\rm int},{\cal M}^2}$} &
\multicolumn{2}{c|}{$\tau_{{\rm int},{\cal S}_2}$} \\[0.1cm]
\hline\hline
%
% What follows is table3.awkout, prepared by rewindow.awk (22 Dec 2003)
%
  4 & 2.1169 & 0.0018 & 2.3697 & 0.0021 & 2.3906 & 0.0022 & 2.3493 & 0.0021 & 2.3830 & 0.0022 \\
  6 & 2.7257 & 0.0026 & 3.0618 & 0.0031 & 3.1098 & 0.0031 & 3.0301 & 0.0031 & 3.0977 & 0.0031 \\
  8 & 3.2298 & 0.0033 & 3.6496 & 0.0040 & 3.7238 & 0.0041 & 3.5959 & 0.0039 & 3.7030 & 0.0041 \\
 12 & 4.0638 & 0.0047 & 4.6314 & 0.0058 & 4.7558 & 0.0060 & 4.5324 & 0.0056 & 4.7187 & 0.0059 \\
 16 & 4.7701 & 0.0027 & 5.4588 & 0.0033 & 5.6303 & 0.0034 & 5.3069 & 0.0031 & 5.5739 & 0.0034 \\
 24 & 5.9567 & 0.0083 & 6.8408 & 0.0102 & 7.0897 & 0.0108 & 6.5789 & 0.0096 & 6.9923 & 0.0105 \\
 32 & 6.9303 & 0.0074 & 7.9625 & 0.0090 & 8.2727 & 0.0096 & 7.5922 & 0.0084 & 8.1362 & 0.0094 \\
 48 & 8.5612 & 0.0073 & 9.8308 & 0.0090 & 10.2429 & 0.0096 & 9.2535 & 0.0083 & 10.0271 & 0.0093 \\
 64 & 9.8955 & 0.0101 & 11.3374 & 0.0124 & 11.8272 & 0.0132 & 10.5715 & 0.0112 & 11.5342 & 0.0127 \\
 96 & 12.1713 & 0.0263 & 13.8983 & 0.0321 & 14.5123 & 0.0343 & 12.7790 & 0.0284 & 14.0740 & 0.0327 \\
128 & 13.9684 & 0.0438 & 15.8963 & 0.0533 & 16.5963 & 0.0567 & 14.4798 & 0.0462 & 16.0373 & 0.0540 \\
192 & 16.8861 & 0.0778 & 19.0981 & 0.0933 & 19.9312 & 0.0996 & 17.0821 & 0.0790 & 19.0814 & 0.0933 \\
256 & 19.3990 & 0.0813 & 21.8291 & 0.0969 & 22.7669 & 0.1034 & 19.4148 & 0.0814 & 21.7656 & 0.0967 \\
[0.1cm]
\hline
\end{tabular}
\caption{
    Autocorrelation times for the Swendsen--Wang algorithm
    at the critical point of the 3-dimensional Ising model.
    For each lattice size ($L$), we report the integrated autocorrelation time
    for the bond occupation ($\tau_{{\rm int},{\cal N}}$),
    the energy  ($\tau_{{\rm int},{\cal E}}$),
    the nearest-neighbor connectivity ($\tau_{{\rm int},{\cal E}^\prime}$),
    the squared magnetization ($\tau_{{\rm int},{\cal M}^2}$),
    and the mean-square cluster size ($\tau_{{\rm int},{\cal S}_2}$).
   The quoted error bar corresponds to one standard deviation
   (i.e.\ confidence level $\approx$ 68\%).
}
\label{table_dynamic_1}
\end{table}

%%
%% TABLE 4: DYNAMIC QUANTITIES [part.II]
%%
\begin{table}%[h]
\centering
%\vspace*{-0.5in} % Move table upwards
%\addtolength{\tabcolsep}{-1.0mm}
%\hspace*{-1.0cm}    % Move table leftwards
%\protect\footnotesize
\begin{tabular}{|r|r@{$\,\pm\,$}r|r@{$\,\pm\,$}r|%
                   r@{$\,\pm\,$}r|r@{$\,\pm\,$}r|}%
\hline  \\[-0.5cm]
\multicolumn{1}{|c|}{$L$}    &
\multicolumn{2}{c|}{$\tau_{{\rm int},{\cal S}_0}$} &
\multicolumn{2}{c|}{$\tau_{{\rm int},{\cal C}_1}$}   &
\multicolumn{2}{c|}{$\tau_{{\rm int},{\cal C}_2}$}   &
\multicolumn{2}{c|}{$\tau_{{\rm int},{\cal C}_3}$} \\[0.1cm]
\hline\hline
%
% What follows is table4.awkout, prepared by rewindow.awk (22 Dec 2003)
%
  4 & 2.0347 & 0.0019 & 2.2994 & 0.0020 & 1.1843 & 0.0010 & 1.4474 & 0.0014 \\
  6 & 2.5473 & 0.0026 & 2.9784 & 0.0029 & 1.3890 & 0.0013 & 1.7324 & 0.0018 \\
  8 & 2.9917 & 0.0033 & 3.5488 & 0.0039 & 1.5639 & 0.0015 & 1.9720 & 0.0022 \\
 12 & 3.7408 & 0.0046 & 4.5025 & 0.0055 & 1.8564 & 0.0020 & 2.3786 & 0.0029 \\
 16 & 4.3847 & 0.0026 & 5.3001 & 0.0031 & 2.0973 & 0.0011 & 2.7099 & 0.0016 \\
 24 & 5.4764 & 0.0082 & 6.6205 & 0.0097 & 2.4807 & 0.0031 & 3.2554 & 0.0046 \\
 32 & 6.3760 & 0.0072 & 7.6876 & 0.0086 & 2.7875 & 0.0026 & 3.6901 & 0.0039 \\
 48 & 7.8999 & 0.0072 & 9.4380 & 0.0085 & 3.2661 & 0.0024 & 4.3843 & 0.0037 \\
 64 & 9.1455 & 0.0100 & 10.8384 & 0.0116 & 3.6451 & 0.0031 & 4.9386 & 0.0049 \\
 96 & 11.2938 & 0.0262 & 13.1816 & 0.0297 & 4.2519 & 0.0074 & 5.8452 & 0.0120 \\
128 & 12.9878 & 0.0439 & 14.9924 & 0.0487 & 4.7240 & 0.0118 & 6.5409 & 0.0193 \\
192 & 15.7728 & 0.0783 & 17.8138 & 0.0842 & 5.4151 & 0.0194 & 7.5908 & 0.0320 \\
256 & 18.1480 & 0.0822 & 20.2905 & 0.0870 & 6.0248 & 0.0193 & 8.5341 & 0.0324 \\
[0.1cm]
\hline
\end{tabular}
\caption{
    Autocorrelation times for the Swendsen--Wang algorithm
    at the critical point of the 3-dimensional Ising model.
    For each lattice size ($L$), we report the integrated autocorrelation time
    for the number of clusters ($\tau_{{\rm int},{\cal S}_0}$)
    and for the sizes of the three largest clusters
    ($\tau_{{\rm int},{\cal C}_i}$ for $i=1,2,3$).
   The quoted error bar corresponds to one standard deviation
   (i.e.\ confidence level $\approx$ 68\%).
}
\label{table_dynamic_2}
\end{table}

\clearpage

%%
%% TABLE 5: fits for critical exponents
%%
\begin{table}
\centering
\begin{tabular}{|l|c|c|c|}%
\hline  %\\[-0.5cm]
%\multicolumn{1}{|c|}{Exponent} & \multicolumn{1}{|c}{numerical}  \\
Exponent & Estimate & $L_{\rm min}$  & $\chi^2$\ (DF, CL)\\
\hline \hline
$z_{{\rm int},{\cal N}}$
              &$0.4745\kk \pm\kk 0.0044$ & 96 &  0.263 (2 DF, 87.7\%)   \\
$z_{{\rm int},{\cal E}}$
              &$0.4599\kk \pm\kk 0.0046$ & 96 &  0.338 (2 DF, 84.4\%) \\
$z_{{\rm int},{\cal E'}}$
              &$0.4588\kk \pm\kk 0.0047$ & 96 &  0.352 (2 DF, 83.8\%)  \\
\hline
$z_{{\rm int},{\cal M}^2}$
              &$0.4245\kk \pm\kk 0.0044$ & 96 &  1.641 (2 DF, 44.0\%)  \\
$z_{{\rm int},{\cal S}_2}$
              &$0.4432\kk \pm\kk 0.0046$ & 96  & 1.126 (2 DF, 57.0\%)  \\
\hline
$z_{{\rm int},{\cal S}_0}$
              &$0.4832\kk \pm\kk 0.0047$ & 96  & 0.082 (2 DF, 96.0\%) \\
$z_{{\rm int},{\cal C}_1}$
              &$0.4384\kk \pm\kk 0.0045$ & 96  & 0.971 (2 DF, 61.5\%) \\
$z_{{\rm int},{\cal C}_2}$
              &$0.3537\kk \pm\kk 0.0034$ &  96 & 2.786 (2 DF, 24.8\%) \\
$z_{{\rm int},{\cal C}_3}$
              &$0.3836\kk \pm\kk 0.0040$ &  96 & 1.975 (2 DF, 37.2\%) \\[1mm]
\hline
\end{tabular}
\caption{
    Numerical estimates for the dynamic critical exponents
    of the 3-dimensional Ising model,
    based on least-squares fits with the specified $L_{\rm min}$.
   The quoted error bar corresponds to one standard deviation
   (i.e.\ confidence level $\approx$ 68\%).
}
\label{table_results}
\end{table}

\clearpage

%%
%% TABLE 6: A and \tau_{exp} (Part 1)
%%
\begin{table}
%\centering
%\vspace*{-0.5in} % Move table upwards
%%\addtolength{\tabcolsep}{-1.0mm}
\hspace*{-4mm}    % Move table leftwards
\protect\small
\begin{tabular}{|r||r|r||r|r||r|r||r|r||r|r|}
\hline  \\[-0.5cm]
\multicolumn{1}{|c||}{$L$}    &
\multicolumn{1}{|c|}{$A_{\scrn}$} &
  \multicolumn{1}{|c||}{$\tau_{{\rm exp},\scrn}$} &
\multicolumn{1}{|c|}{$A_{\scre}$} &
  \multicolumn{1}{|c||}{$\tau_{{\rm exp},\scre}$} &
\multicolumn{1}{|c|}{$A_{\scre'}$} &
  \multicolumn{1}{|c||}{$\tau_{{\rm exp},\scre'}$} &
\multicolumn{1}{|c|}{$A_{\scrm^2}$} &
  \multicolumn{1}{|c||}{$\tau_{{\rm exp},\scrm^2}$} &
\multicolumn{1}{|c|}{$A_{\scrs_2}$} &
  \multicolumn{1}{|c|}{$\tau_{{\rm exp},\scrs_2}$}  \\[0.1cm]
\hline\hline
4 & 0.8496 & 2.3588 & 0.9843 & 2.3628 & 0.9975 & 2.3611 & 0.9738 & 2.3604 & 0.9937 & 2.3602  \\
6 & 0.8372 & 3.1071 & 0.9704 & 3.1059 & 0.9895 & 3.1080 & 0.9561 & 3.1063 & 0.9845 & 3.1078  \\
8 & 0.8284 & 3.7396 & 0.9622 & 3.7414 & 0.9875 & 3.7407 & 0.9440 & 3.7411 & 0.9810 & 3.7397  \\
12 & 0.8157 & 4.8034 & 0.9524 & 4.8069 & 0.9844 & 4.8060 & 0.9275 & 4.8062 & 0.9748 & 4.8065  \\
16 & 0.8028 & 5.7303 & 0.9406 & 5.7284 & 0.9755 & 5.7308 & 0.9090 & 5.7305 & 0.9638 & 5.7323  \\
24 & 0.7919 & 7.2722 & 0.9283 & 7.2692 & 0.9682 & 7.2687 & 0.8899 & 7.2575 & 0.9548 & 7.2621  \\
32 & 0.7857 & 8.5472 & 0.9196 & 8.5443 & 0.9615 & 8.5429 & 0.8728 & 8.5374 & 0.9451 & 8.5379  \\
48 & 0.7827 & 10.6314 & 0.9112 & 10.6380 & 0.9548 & 10.6364 & 0.8532 & 10.6385 & 0.9333 & 10.6400  \\
64 & 0.7785 & 12.3751 & 0.9037 & 12.3777 & 0.9473 & 12.3835 & 0.8365 & 12.4059 & 0.9218 & 12.4023  \\
96 & 0.7709 & 15.3193 & 0.8879 & 15.3373 & 0.9326 & 15.3276 & 0.8126 & 15.3481 & 0.9041 & 15.3353  \\
128 & 0.7748 & 17.6210 & 0.8884 & 17.6409 & 0.9316 & 17.6424 & 0.8043 & 17.6836 & 0.8987 & 17.6729  \\
192 & 0.7782 & 21.2902 & 0.8867 & 21.2979 & 0.9289 & 21.3083 & 0.7949 & 21.2413 & 0.8933 & 21.2720  \\
256 & 0.7679 & 24.6645 & 0.8725 & 24.6212 & 0.9133 & 24.6273 & 0.7728 & 24.6522 & 0.8729 & 24.6519  \\
\hline
\end{tabular}
\caption{
    Estimates of the amplitude $A_{\scro}$
    and the exponential autocorrelation time $\tau_{{\rm exp},\scro}$
    for the observables $\scrn$, $\scre$, $\scre'$, $\scrm^2$ and $\scrs_2$.
}
\label{table_exponential_1}
\end{table}

%%
%% TABLE 7: A and \tau_{exp} (Part 2)
%%
\begin{table}
\centering
%\vspace*{-0.5in} % Move table upwards
%%\addtolength{\tabcolsep}{-1.0mm}
%%\hspace*{-3mm}    % Move table leftwards
\protect\small
\begin{tabular}{|r||r|r||r|r||r|r||r|r||}
\hline  \\[-0.5cm]
\multicolumn{1}{|c||}{$L$}    &
\multicolumn{1}{|c|}{$A_{\scrs_0}$} &
  \multicolumn{1}{|c||}{$\tau_{{\rm exp},\scrs_0}$} &
\multicolumn{1}{|c|}{$A_{\scrc_1}$} &
  \multicolumn{1}{|c||}{$\tau_{{\rm exp},\scrc_1}$} &
\multicolumn{1}{|c|}{$A_{\scrc_2}$} &
  \multicolumn{1}{|c||}{$\tau_{{\rm exp},\scrc_2}$} &
\multicolumn{1}{|c|}{$A_{\scrc_3}$} &
  \multicolumn{1}{|c|}{$\tau_{{\rm exp},\scrc_3}$}  \\[0.1cm]
\hline\hline
4 & 0.7991 & 2.3582 & 0.9479 & 2.3499 & 0.3298 & 2.3284 & 0.4858 & 2.3399  \\
6 & 0.7631 & 3.1055 & 0.9342 & 3.0971 & 0.3028 & 3.0851 & 0.4494 & 3.0761  \\
8 & 0.7500 & 3.7358 & 0.9297 & 3.7252 & 0.2931 & 3.7079 & 0.4318 & 3.7036  \\
12 & 0.7361 & 4.7983 & 0.9207 & 4.7843 & 0.2781 & 4.7952 & 0.4108 & 4.7733  \\
16 & 0.7248 & 5.7269 & 0.9078 & 5.7042 & 0.2698 & 5.7093 & 0.3970 & 5.6844  \\
24 & 0.7165 & 7.2712 & 0.8936 & 7.2358 & 0.2612 & 7.2010 & 0.3809 & 7.2192  \\
32 & 0.7131 & 8.5471 & 0.8838 & 8.4986 & 0.2520 & 8.5026 & 0.3710 & 8.4798  \\
48 & 0.7139 & 10.6296 & 0.8704 & 10.5925 & 0.2425 & 10.6074 & 0.3582 & 10.5656  \\
64 & 0.7125 & 12.3728 & 0.8585 & 12.3428 & 0.2353 & 12.3740 & 0.3471 & 12.3565  \\
96 & 0.7085 & 15.3250 & 0.8406 & 15.2511 & 0.2250 & 15.3425 & 0.3358 & 15.2290  \\
128 & 0.7154 & 17.6157 & 0.8335 & 17.5967 & 0.2214 & 17.6720 & 0.3301 & 17.5485  \\
192 & 0.7218 & 21.2930 & 0.8265 & 21.2023 & 0.2146 & 21.3083 & 0.3242 & 21.0661  \\
256 & 0.7139 & 24.6895 & 0.8078 & 24.5465 & 0.2076 & 24.7020 & 0.3105 & 24.6323  \\
\hline
\end{tabular}
\caption{
    Estimates of the amplitude $A_{\scro}$
    and the exponential autocorrelation time $\tau_{{\rm exp},\scro}$.
    for the observables $\scrs_0$, $\scrc_1$, $\scrc_2$ and $\scrc_3$.
}
\label{table_exponential_2}
\end{table}

\clearpage

%%
%% TABLE 8: A and \tau_{int}/\tau_{exp} (Part 1)
%%
\begin{table}
%\centering
%\vspace*{-0.5in} % Move table upwards
%%\addtolength{\tabcolsep}{-1.0mm}
\hspace*{-0mm}    % Move table leftwards
\protect\small
\begin{tabular}{|r||r|r||r|r||r|r||r|r||r|r|}
\hline  \\[-0.5cm]
\multicolumn{1}{|c||}{$L$}    &
\multicolumn{1}{|c|}{$A_{\scrn}$} &
  \multicolumn{1}{|c||}{$R_{\scrn}$} &
\multicolumn{1}{|c|}{$A_{\scre}$} &
  \multicolumn{1}{|c||}{$R_{\scre}$} &
\multicolumn{1}{|c|}{$A_{\scre'}$} &
  \multicolumn{1}{|c||}{$R_{\scre'}$} &
\multicolumn{1}{|c|}{$A_{\scrm^2}$} &
  \multicolumn{1}{|c||}{$R_{\scrm^2}$} &
\multicolumn{1}{|c|}{$A_{\scrs_2}$} &
  \multicolumn{1}{|c|}{$R_{\scrs_2}$}  \\[0.1cm]
\hline\hline
  4 & 0.8496 & 0.8834 & 0.9843 & 0.9889 & 0.9975 & 0.9976 & 0.9738 & 0.9804 & 0.9937 & 0.9945   \\
  6 & 0.8372 & 0.8695 & 0.9704 & 0.9767 & 0.9895 & 0.9920 & 0.9561 & 0.9666 & 0.9845 & 0.9882   \\
  8 & 0.8284 & 0.8583 & 0.9622 & 0.9699 & 0.9875 & 0.9896 & 0.9440 & 0.9556 & 0.9810 & 0.9841   \\
 12 & 0.8157 & 0.8425 & 0.9524 & 0.9602 & 0.9844 & 0.9860 & 0.9275 & 0.9397 & 0.9748 & 0.9783   \\
 16 & 0.8028 & 0.8303 & 0.9406 & 0.9501 & 0.9755 & 0.9800 & 0.9090 & 0.9237 & 0.9638 & 0.9702   \\
 24 & 0.7919 & 0.8182 & 0.9283 & 0.9397 & 0.9682 & 0.9738 & 0.8899 & 0.9037 & 0.9548 & 0.9605   \\
 32 & 0.7857 & 0.8103 & 0.9196 & 0.9310 & 0.9615 & 0.9673 & 0.8728 & 0.8877 & 0.9451 & 0.9513   \\
 48 & 0.7827 & 0.8043 & 0.9112 & 0.9236 & 0.9548 & 0.9623 & 0.8532 & 0.8693 & 0.9333 & 0.9420   \\
 64 & 0.7785 & 0.7987 & 0.9037 & 0.9150 & 0.9473 & 0.9546 & 0.8365 & 0.8532 & 0.9218 & 0.9309   \\
 96 & 0.7709 & 0.7938 & 0.8879 & 0.9064 & 0.9326 & 0.9465 & 0.8126 & 0.8334 & 0.9041 & 0.9179   \\
128 & 0.7748 & 0.7915 & 0.8884 & 0.9008 & 0.9316 & 0.9405 & 0.8043 & 0.8205 & 0.8987 & 0.9088   \\
192 & 0.7782 & 0.7923 & 0.8867 & 0.8961 & 0.9289 & 0.9352 & 0.7949 & 0.8015 & 0.8933 & 0.8953   \\
256 & 0.7679 & 0.7876 & 0.8725 & 0.8863 & 0.9133 & 0.9243 & 0.7728 & 0.7882 & 0.8729 & 0.8837   \\
\hline
\end{tabular}
\caption{
    Estimates of the amplitude $A_{\scro}$
    and the ratio $R_{\scro} = \tau_{{\rm int},{\cal O}}/\bar{\tau}_{\rm exp}$
    for the observables $\scrn$, $\scre$, $\scre'$, $\scrm^2$ and $\scrs_2$.
}
\label{table_ratios_1}
\end{table}

%%
%% TABLE 9: A and \tau_{int}/\tau_{exp} (Part 2)
%%
\begin{table}
\centering
%\vspace*{-0.5in} % Move table upwards
%%\addtolength{\tabcolsep}{-1.0mm}
%%\hspace*{-4mm}    % Move table leftwards
\protect\small
\begin{tabular}{|r||r|r||r|r||r|r||r|r|}
\hline  \\[-0.5cm]
\multicolumn{1}{|c||}{$L$}    &
\multicolumn{1}{|c|}{$A_{\scrs_0}$} &
  \multicolumn{1}{|c||}{$R_{\scrs_0}$} &
\multicolumn{1}{|c|}{$A_{\scrc_1}$} &
  \multicolumn{1}{|c||}{$R_{\scrc_1}$} &
\multicolumn{1}{|c|}{$A_{\scrc_2}$} &
  \multicolumn{1}{|c||}{$R_{\scrc_2}$} &
\multicolumn{1}{|c|}{$A_{\scrc_3}$} &
  \multicolumn{1}{|c|}{$R_{\scrc_3}$}  \\[0.1cm]
\hline\hline
  4 & 0.7991 & 0.8491 & 0.9479 & 0.9596 & 0.3298 & 0.4942 & 0.4858 & 0.6040  \\
  6 & 0.7631 & 0.8126 & 0.9342 & 0.9501 & 0.3028 & 0.4431 & 0.4494 & 0.5527  \\
  8 & 0.7500 & 0.7950 & 0.9297 & 0.9431 & 0.2931 & 0.4156 & 0.4318 & 0.5241  \\
 12 & 0.7361 & 0.7756 & 0.9207 & 0.9335 & 0.2781 & 0.3849 & 0.4108 & 0.4931  \\
 16 & 0.7248 & 0.7632 & 0.9078 & 0.9225 & 0.2697 & 0.3650 & 0.3970 & 0.4717  \\
 24 & 0.7165 & 0.7522 & 0.8936 & 0.9094 & 0.2612 & 0.3407 & 0.3809 & 0.4472  \\
 32 & 0.7131 & 0.7455 & 0.8838 & 0.8989 & 0.2520 & 0.3259 & 0.3710 & 0.4315  \\
 48 & 0.7139 & 0.7422 & 0.8704 & 0.8867 & 0.2425 & 0.3068 & 0.3582 & 0.4119  \\
 64 & 0.7125 & 0.7381 & 0.8585 & 0.8748 & 0.2353 & 0.2942 & 0.3471 & 0.3986  \\
 96 & 0.7085 & 0.7366 & 0.8406 & 0.8597 & 0.2250 & 0.2773 & 0.3358 & 0.3812  \\
128 & 0.7154 & 0.7360 & 0.8335 & 0.8496 & 0.2214 & 0.2677 & 0.3301 & 0.3706  \\
192 & 0.7218 & 0.7401 & 0.8265 & 0.8358 & 0.2146 & 0.2541 & 0.3242 & 0.3562  \\
256 & 0.7139 & 0.7368 & 0.8078 & 0.8238 & 0.2076 & 0.2446 & 0.3105 & 0.3465  \\
\hline
\end{tabular}
\caption{
    Estimates of the amplitude $A_{\scro}$
    and the ratio $R_{\scro} = \tau_{{\rm int},{\cal O}}/\bar{\tau}_{\rm exp}$
    for the observables $\scrs_0$, $\scrc_1$, $\scrc_2$ and $\scrc_3$.
}
\label{table_ratios_2}
\end{table}

\clearpage

%%
%% TABLE 10: exponents and theoretical predictions
%%
\begin{table}
%\centering
\hspace*{-1.3cm}    % Move table leftwards
\addtolength{\tabcolsep}{-0.5mm}
\protect\small
\begin{tabular}{|c||c|c|c|c||c|c|c|}
\hline  %\\[-0.5cm]
Model & $\alpha/\nu$ & $\beta/\nu$ & $\gamma/\nu$ & $z_{\rm G}$ &
  $z_{\rm KRT}^{\rm (lower)}$ & $z_{\rm SW}$ & $z_{\rm KRT}^{\rm (upper)}$  \\
\hline \hline
$d=1$, all $q$ &  --- &  0  &  1  &
    2 &
    0 &
    0 &
    0 \\
\hline
$d=2$ Ising  &  0 $(\times\log)$  &  1/8  &  7/4  &
    2.16(2) &
    0.29(2) &
    0.222(7) \protect\cite{Salas_Sokal_Ising_v1} &
    0.41(2) \\
\hline
$d=2$, $q=3$ &  2/5  &  2/15   &  26/15  &
    2.19(2) &
    0.33(2) &
    0.514(6) \protect\cite{Salas_Sokal_Potts3} &
    0.46(2) \\
\hline
$d=2$, $q=4$ &
    1 $(\times\log^{-3/2})$  &
    1/8 $(\times\log^{1/16})$  &
    7/4 $(\times\log^{-1/8})$  &
    2.25(6) &
    0.38(6) &
    1 $(\times\log^{??})$ \protect\cite{Salas_Sokal_Potts4} &
    0.50(6) \\
\hline
$d=3$ Ising  &
    0.1756(25) &
    0.5183(4) &
    1.9634(8) &
    2.04(4) &
    0.46(4) &
    0.46(3) [this work] &
    0.73(4) \\
\hline
$d=4$ Ising  &
    0 $(\times\log^{1/3})$  &
    1 $(\times\log^{1/6})$  &
    2 &
    2 &
    2/3 &
    1 $(\times\log^{??})$ &
    1 \\
\hline
\end{tabular}
\caption{
    Dynamic critical exponent $z_{{\rm int},{\cal E}'}$
    of the Swendsen--Wang algorithm for various Potts models,
    compared to the exponents arising in the theoretical frameworks of
    Li and Sokal \protect\cite{Li_Sokal},
    Klein, Ray and Tamayo \protect\cite{Klein_89},
    and Coddington and Baillie \protect\cite{Coddington_92}.
    Static critical exponents $\alpha/\nu$, $\beta/\nu$ and $\gamma/\nu$
    are exact values for the $d=1$ and $d=2$ models
    \cite{Baxter,Nienhuis_84,DiFrancesco_97,Salas_Sokal_Potts4}
    and for $d=4$ Ising \cite{Brezin_76},
    %%
    %% From Brezin_76, section VIII.B for d=4 n-vector model:
    %%
    %%   \chi \sim \xi^2 without logarithms (because \eta starts at \epsilon^2)
    %%
    %%   \chi  \sim  t^{-1} (\log t)^{(n+2)/(n+8)}
    %%   M     \sim  (-t)^{1/2} (\log t)^{3/(n+8)}
    %%   C_H   \sim  t^{-1} (\log t)^{(4-n)/(n+8)}
    %%
    %% So we can deduce logarithms in everything as a function of \xi.
    %%
    and are the best currently available numerical estimates
    for $d=3$ Ising \cite{Hasenbusch_99}.
    Exponent $z_{\rm G}$ for Glauber dynamics
    is taken from \cite[Table 1]{daSilva_02} for the $d=2$ models
    (see also \cite{Munkel_93,Linke_95,Stauffer_97,Li_97,Ying_01}),
    from \cite{Wansleben_91,Heuer_92,Munkel_93,Grassberger_95,%
Stauffer_97,Zheng_98,Jaster_99,Ito_00,Calabrese_03} for $d=3$ Ising,
    and from \cite{Hohenberg_77} for $d=4$ Ising.
    %%  z = 2 + O(\epsilon^2) in d=4-\epsilon, so no logs in d=4
}
\label{table_discussion}
\end{table}

\clearpage

%%%%%%%%%%%% START OF FIGURES %%%%%%%%%%%%%

%
% FIGURE 1
%
% Least-squares fit to \tau_{int,E'}
%
%

\clearpage

%
% FIGURE 2
%
% autocorrelation function of E' and C2 for L=64
%
%

\clearpage

%
% FIGURE 3
%
% Least-squares fit to \tau_{exp}
%
%

\clearpage

%
% FIGURE 4
%
% autocorrelation function of E' for all L
%
%

\clearpage

%
% FIGURE 5
%
% Least-squares fit to \tau_{int,E'} C_1 / L^d
%
%

\clearpage

%%\doublespace
%%\listoffigures

%%\clearpage

%%\doublespace
%%\listoftables

%%\clearpage

\end{document}